# Electro-Oxidation of Ni42 Steel: A highly Active Bifunctional Electrocatalyst

Helmut Schäfer*[a], Daniel M. Chevrier [b], Peng Zhang [b], Johannes Stangl [c], Klaus Müller-Buschbaum [c], Jörg D. Hardege[d], Karsten Kuepper[ e, f], Joachim Wollschläger [ e, f] , Ulrich Krupp[g], Simon Dühnen [a], Martin Steinhart [a], Lorenz Walder [a], Shamaila Sadaf [a], and Mercedes Schmidt[a]

[a]*Institute of Chemistry of New Materials and Center of Physics and Chemistry of New Materials, Universität Osnabrück, Barbarastrasse 7, 49076 Osnabrück, Germany*

[b] *Department of Chemistry, Dalhousie University, Halifax, Nova Scotia, Canada B3H 4J3*

[c]*University of Würzburg, Institute of Inorganic Chemistry Julius-Maximilians-Universität Würzburg Am Hubland, 97074 Würzburg, Germany*

[d] *School of Biological, Biomedical and Environmental Sciences, University of Hull, Cottingham Road, Hull, HU6 7RX, United Kingdom*

[e] *Department of Physics, Universität Osnabrück, Barbaraßtraße 7, 49069 Osnabrück, Germany*

[f] *Center of Physics and Chemistry of New Materials, Universität Osnabrück, Barbaraßtraße 7, 49069 Osnabrück, Germany*

[g] *Institute of Materials Design and Structural Integrity University of Applied Sciences Osnabrück, Albrechtstraße 30, 49076 Osnabrück, Germany*

Janus type Water-Splitting Catalysts have attracted highest attention as a tool of choice for solar to fuel conversion. AISI Ni 42 steel was upon harsh anodization converted in a bifunctional electrocatalyst. Oxygen evolution reaction- (OER) and hydrogen evolution reaction (HER) are highly efficiently and steadfast catalyzed at pH 7, 13, 14, 14.6 (OER) respectively at pH 0, 1, 13, 14, 14.6 (HER). The current density taken from long-term OER measurements in pH 7 buffer solution upon the electro activated steel at 491 mV overpotential ($\eta$) was around 4 times higher (4 mA/cm$^2$) in comparison with recently developed OER electrocatalysts. The very strong voltage-current behavior of the catalyst shown in OER polarization experiments at both pH 7 and at pH 13 were even superior to those known for IrO$_2$-RuO$_2$. No degradation of the catalyst was detected even when conditions close to standard industrial operations were applied to the catalyst. A stable Ni-, Fe- oxide based passivating layer sufficiently protected the bare metal for further oxidation. Quantitative charge to oxygen- (OER) and charge to hydrogen (HER) conversion was confirmed. High resolution XPS spectra showed that most likely γ–NiO(OH) and FeO(OH) are the catalytic active OER and NiO is the catalytic active HER species.

# 1. Introduction

The limited availability of primary forms of energy including non-renewable fossil based energy sources such as oil and gas is increasingly forcing engineers and scientists to develop techniques to extract electric energy from renewable energy sources [1, 2, 3, 4, 5, 6]. Electrochemically initiated water splitting allows the conversion of electricity into fuel- hydrogen plus oxygen, with which a fuel cell can be operated [7, 8, 9, 10, 11, 12,]. Thus photo synthesis, i.e. the conversion of light energy into "chemical energy" can be artificially realized via the combination of e.g. light driven solar cells plus a water electrolysis cell [13]. It is well known that the effectiveness of this attractive solar to fuel conversion route is severely restricted by the high overpotentials caused by commonly used electrode materials on the anode side [14, 15]. This is particularly true when the electrochemical cleavage of more or less untreated water is performed, when the splitting procedure is carried out at neutral pH value [14]. Ideally electrode materials that have proved to be highly attractive towards electrochemically driven splitting of water in alkaline regime, should also show superior OER characteristics in neutral regimes ensured by potentials that are stable over long operating times and do not differ much from the reversible $H_2O/O_2$ potential (1.228 V *vs.* RHE at 298.15 K). Particularly OER electrocatalysts that function under neutral conditions are promising in as much as they could form part of a concept for seawater-splitting using renewable energy sources. In addition to multiple pH suitability of OER electrocatalysts, such state of the art electrocatalysts will definitely derive further benefits from their multi-functionality, i.e. should be an ideal tool to catalyse both OER plus HER.

To date the development of cheap, earth abundant electrocatalysts exhibiting appropriate characteristics regarding catalytic activity and stability whilst initiating HER plus OER over a broad pH range is still considered to be very challenging.

Noble metals like Pt, Ir, Ru, Au or noble metal containing compounds like $IrO_2$-$RuO_2$ , $IrNiO_x$ , $RuO_2$-NiO are famous for their comparatively low overpotential when used as an OER electrocatalyst in both alkaline and neutral regime [16, 17, 18, 19, 20]. The often claimed high costs of iridium and ruthenium can, at present not be considered as a serious obstacle taken into consideration the fact that commercially available $IrO_2$-$RuO_2$ based catalysts made off micrometre thick, PVD sputtered $IrO_2$-$RuO_2$ layers deposited on titanium support material are rarely far more expensive than 6 \$/100 cm$^2$. However it is hard to deny that for widespread use in catalysis the noble metals, iridium as well as ruthenium need to be replaced because of their scarcity in Earth's crust. The electrocatalytic OER properties of Copper oxide based films were recently improved significantly by Sun et al. [21]. For decades Nickel metal as well as Ni containing compounds are known for their sophisticated properties in terms of electrocatalysis [22]. The high OER performance of known Ni containing alloys and oxides is to the best of our knowledge at least up to some extent restricted to alkaline media, i.e. at higher pH regions [23, 24, 25]. Thus Chatenet et al. reported on very good electrocatalytic OER properties of AISI 316 steel under alkaline conditions [26]. Different metals and steel types are since decades in the focus of interest for electrochemical applications [27, 28, 29]. Chlorine oxidized AISI 304 steel (Cr-Ni stainless steel) was found to be a mediocre OER electrocatalyst requiring 500 mV overpotential to ensure 0.65 mA cm$^{-2}$ current density in pH 7 corrected 0.1 M $K_2HPO_4$/$KH_2PO_4$ solution [23]. Generally the determined overpotentials for OER on recently developed materials in pH 7 media are relatively close to 500 mV at 1 mA cm$^{-2}$ current density [30]. An overpotential of 610 mV for the onset of OER at pH 6.9 was reported for $ZrS_3$ nanosheets [31]. Wu et al. [30] reported a 480 mV overpotential at stable current density of 1 mA/cm$^2$ for iron based thin films. Ramirez et. al. evaluated $MnO_x$, $Mn_2O_3$, and $Mn_3O_4$ electrodeposited films for OER in alkaline and neutral media and under optimal synthesis condition the films exhibited sufficient water splitting at pH 7 (470 mV overpotential at 1 mA/cm$^2$ current density)[32]. Very recently we reported on surface oxidized mild steel S 235 as a prospective electrode material for the anodic splitting of water under neutral conditions and determined for the OER in 0.1 M phosphate (pH 7) buffer solution $\eta$=462 mV at 1 mA/cm² current density [33]. We suggested manganese- and iron-oxide species as the potential catalytic active species on the periphery of the chlorinated steel [33]. Electrodeposited FeO(OH) [34, 35] besides cobalt-phosphate compounds [36, 37] are among the known non noble compounds for which slightly better OER activity (down to 410 mV overpotential at 1 mA/cm$^2$) and acceptable stability during water splitting at neutral pH has been demonstrated [37]. However even these state of the art electrocatalysts exhibit at pH 7 a OER performance far below that of the PSII tetrameric Mn clusters which is able to sufficient split water under neutral conditions at overpotentials around 160 mV [38, 39, 40, 41]. Thus under alkaline conditions non-noble metal-based OER electrocatalysts efficiently split water at overpotentials in the range of 200 mV [42, 43, 44, 45, 46] and show in addition high catalytic stability. In contrast the evolution of non-noble metal-based OER electrocatalysts exhibiting similar steady state performance in neutral medium is not nearly as far. In addition multiple pH characteristics of earth abundant materials, i.e. the suitability to catalyze OER over a wide pH range on benchmark level which would pave the way for the widespread use has to the best of our knowledge not been shown so far. Electrocatalytically initiated hydrogen evolution is considered to be less demanding than the oxygen evolution reaction. High catalytic activity for HER with overpotentials far below 200 mV at 10 mA/cm$^2$ current density under alkaline or acidic conditions was proven for a large number of materials [47, 48, 49, 50, 51, 52, 53]. However, very rarely groups succeeded in designing an electrocatalyst that shows both, substantial OER + HER activity [53, 54]. Electrocatalysts that are suitable to support OER as well as HER in the same media can be utilized for the bipolar electrolysers of water electrolysis [55]. Conductive plates which are placed in an electrolyte between two „outer" electrodes connected to the positive and the negative pole of a power source will carry opposite charges on their periphery transforming the electrodes to so-called bipolar electrodes. A bipolar cell configuration substantially simplifies the design of a water-splitting device as both, OER and HER are catalysed upon each of these bipolar electrodes and only both "outer" electrodes but not the central electrodes require a direct power feed.

We herewith show that a surface modification of AISI Ni42 steel converts it to a highly efficient OER electrocatalyst working under neutral- and alkaline conditions (pH 7, 13, 14 and 14,6). In addition HER was initiated at pH 0, 1, 13, 14 and 14.6 upon those steel samples at sufficient low overpotentials leading to a remarkably stable HER current.

## 2. Results and Discussion

**2.1 OER properties in neutral medium.** Figure 1 represents a comparison of the electrochemical OER properties of AISI Ni 42 which had been anodized prior to electrocatalysis in 7.2 M NaOH at 2000 mA/cm$^2$ for 300 min, henceforth referred as sample Ni42-300, with untreated AISI Ni 42, designated as sample Ni42 in pH 7 phosphate buffer solution. The significant enhancement of the current to voltage ratio determined in polarization experiments in neutral medium of Ni 42 alloy upon electro activation can be taken from both, the non-steady state- (cyclic voltammetry; Figures 1a-c) and the steady state- (chronopotentiometry) measurements (Figure 1d). The current density of the OER initiated in pH 7 medium upon sample Ni42-300 within the voltage range 1-1.9 V *versus* RHE reaches its maximum of ~ 12 mA/cm$^2$ at a potential of 1.9 V *versus* RHE (Figure 1a) whereas untreated Ni-42 ensured at 1.9 V *versus* RHE less than 2 mA/cm$^2$ (Figure 1b). The cyclic voltammogram of Ni42 (Figure 1b) showed no signs of Ni existing on the surface of the alloy. Remarkably, neither was the characteristic Ni(II)-Ni(III) oxidation wave obtained, nor any peak in the cathodic part of the CV could be assigned to typical Ni(III)-Ni(II) transition. In contrast, a significant cathodic peak at a potential of ~ 1.30 V *versus* RHE exists in the CV of sample Ni 42-300 (Figure 1a) that can be clearly assigned to the Ni(III)-Ni(II) reduction. From OER on pure Ni electrodes is known that there exists a direct proportionality between the amount of active material based on the Ni(II)/Ni(III) redox system and the charge capacity, Q, calculated by integrating the cathodic voltammetric sweep between the uppermost limit and ca. 1 V *vs.* RHE [56, 57]. Such an analysis performed on the CV given in Figure 1a between ~1.55 V and 1.04 V *versus* RHE delivers Q=11.55 mC/cm$^2$ (Ni 42-300) determined in 0.1 M K$_2$HPO$_4$/KH$_2$PO$_4$ pH 7 solution at 20mV/s sweep rate (Figure 1c). This impressively implies a high amount of electrocatalytic active material. In contrary the CV of untreated ASIS Ni42 showed no cathodic voltammetric sweep suggesting a capacity of 0 (Figure 1b). We recently reported on electro-activated AISI304 steel as a potential OER electrocatalyst in alkaline regimes, i.e. in 0.1 M KOH and found under optimized surface modification conditions for the integrated Q of the corresponding CV curve a value of 27.57 mC/cm$^2$ at 20mV/s sweep rate [45]. However Ni based compounds have rarely been considered as an anode for water-splitting in neutral medium, and to the best of our knowledge the charge capacity Q as a scale for the amount of catalytic active mass referring to the Ni(II)-Ni(III) redox system has not been determined in pH 7 medium.

Chronopotentiometry data demonstrate the very high stability of sample Ni42-300 during electrocatalysis under working conditions (0.1 M KH$_2$PO$_4$/K$_2$HPO$_4$ pH 7 solution; 4 mA/cm$^2$ current density) see Figure 1d, gray curve and in addition underpin the high OER performance of sample Ni42-300 derived from non-steady state measurements (Figure 1a). The potential required to ensure 4 mA/cm$^2$ OER current density in pH 7 phosphate buffer did not vary substantially in the course of 40000 s and amounted to 1.719 V *vs.* RHE. This corresponds to 491 mV of overpotential for the OER (Figure 1d, black curve) in 0.1 M K$_2$HPO$_4$/KH$_2$PO$_4$. A similar outcome was achieved from a 450 ks chronopotentiometry plot ($\eta$= 502 mV at 4 mA/cm$^2$; Figure S1 black curve). These data represent an outstanding OER activity under neutral condition and represent a significant enhancement when compared with the OER characteristics of sample Ni42 (Figure 1d, red curve). On untreated Ni42 steel the required overpotential for OER to guarantee 4 mA/cm$^2$ current density in 0.1 M K$_2$HPO$_4$/KH$_2$PO$_4$ was positively shifted by ~ 180 mV when compared to sample Ni42-300 ($\eta$= 668 mV; Figures 1d, gray curve plus black curve). This impressively confirms the increase of OER performance of Ni42 steel at pH 7 upon this type of surface oxidation. To date the highest electrocatalytic activity for OER on steel surfaces under neutral conditions was reported for surface activated mild steel S235 exhibiting $\eta$= 462 mV at 1 mA/cm$^2$ in pH 7 buffered solution [33]. In general most of the established OER electrocatalysts show at similar overpotentials of around 500 mV a OER current density that is approximately four times lower (1 mA/cm$^2$) [31, 30, 34, 35, 36, 37]. In polarization measurements Co$_3$O$_4$ nanowire arrays [58] developed by He *et al.* showed an unusual strong voltage-current behavior at pH 7.2. However no steady state measurements performed at pH 7.2 were shown and Faraday efficiency was not determined, which makes it at least difficult to conclusively assess the overpotential for the OER at a predetermined current density. Very recently the OER performance of Graphene-$Co_3O_4$ nanocomposites have been investigated under neutral conditions and also exhibited similarly low overpotentials for the OER derived from cyclic voltammograms (498 mV at 10 mA cm$^{-2}$) [59]. Unfortunately no long-term voltage-current behavior was shown and the release of oxygen was not quantified [59].

The release of gas bubbles visible to the eye on the surface of sample Ni42-300 was observed throughout the chronopotentiometry measurements. Our electro-activation procedure creates on the AISI Ni 42 steel a passivating oxide layer which sufficiently protects the metal matrix below the layer against further oxidation. Cross sectional analysis (vertical plane imaging) of samples was performed by dual beam FIB–SEM (focused ion beam) technique in order to estimate the thickness of the oxide layer (Figure S2). The thickness was around 4.7 μm. To confirm the assumption that the current determined upon electrochemical measurements is due to oxygen evolution we determined the Faradaic Efficiency by direct fluorescence-based sensing of the evolved oxygen (Figure 2a) during chronopotentiometry at constant current density of 2 mA/cm$^2$ in 0.1 M pH 7 phosphate solution (Figure 2b). The curve course of the dissolved oxygen (Figure 2a) as a function of time (dotted curve) shows a good agreement with the theoretically possible increase of dissolved oxygen on the basis of a 100% charge to oxygen conversion (100% Faradaic efficiency; Figure 2a, red line). The Faradaic efficiency of the OER upon sample Ni 42-300 (in 0.1 M $K_2HPO_4$/$KH_2PO_4$ at 2 mA/cm$^2$ and at 470 mV overpotential for the OER) was found to be 99.4 % after 4000 s running time (Figure 2a), proving the overall brilliant electrocatalytic oxygen evolution properties (see Experimental section for details). The Faradaic Efficiency regarding OER of mild steel S235 at the same current density was significantly lower and amounted to only 67% after 3000 s of chronopotentiometry in 0.1 M KOH [33]. The Faradaic efficiency of a variety of OER electrocatalysts under neutral conditions has been investigated by several groups and in most cases exhibited significantly lower values than in the current study. Manganese oxides based OER electrocatalysts as a model system for the oxygen evolving complex of photosystem II were frequently studied and based on head space measurements the Faradaic efficiency amounted to 78% after 90 min of OER in pH controlled $Na_2SO_4$ [60]. The charge to oxygen conversion rates of recently developed full water splitting electrocatalysts determined in alkaline solution ranged between 95.8 and 100% [109, 110, 112]. A comparison of the OER properties of Ni42-300 at pH 7 with the corresponding characteristics of noble metal containing catalysts is a sine qua for an in-depth evaluation of the electrocatalytic characteristics of our material. We chose commercially available $IrO_2$-$RuO_2$ sputtered on titanium as reference sample (sample $IrO_2$-$RuO_2$) for OER activity and stability at pH 7 and pH 13. A direct comparison of the OER performance of sample Ni42-300 and $IrO_2$-$RuO_2$ can be derived from Figure 3 and reveals the non-steady state (3a) as well as the steady state (3b) voltage-current behaviors regarding OER based polarization experiments. In the cyclic voltammograms shown in Figure 3a it can be seen that astonishingly only at very low overpotentials (< 200 mV) does $IrO_2$-$RuO_2$ (Figure 3a, green curve) show slightly higher current densities in comparison with the electro-oxidized Ni42 (Figure 3a, black curve). Over the entire range of overpotentials, typically important regarding OER in neutral regime (200 mV > η < 700 mV) sample Ni42-300 exhibited superior electrocatalytic OER activity when compared to $IrO_2$-$RuO_2$ (Figure 3a). Especially at elevated potentials (E ≥ 1.7 V *vs.* RHE) sample $IrO_2$-$RuO_2$ loses performance significantly (Figure 3a, gray curve). In comparison with the outcome of our earlier studies [23, 45, 33] with respect to the overall electrocatalytic OER specifications of surface treated steel under neutral conditions, our current results present a significant improvement. Whereas surface activated AISI 304 steel showed at pH 13 superior electrocatalytic OER properties when compared to $IrO_2$-$RuO_2$ [45], our group failed until recently to significantly improve the OER properties of steel upon surface oxidation to a level that makes it at least nearly competitive to $IrO_2$-$RuO_2$ regarding water-splitting properties under neutral conditions [23, 33].

Tafel plots (Figure 3b) reflect the findings based on non-steady state electrochemical measurements made. Whereas in the lower overpotential region (< 500 mV) $IrO_2$-$RuO_2$ (Figure 3b, green squares) exhibited substantial higher upon OER initiated current density at a defined potential when compared to sample Ni42-300 (Figure 3b, black squares), this changes at an overpotential of ~ 550 mV. Sample Ni42-300 reached (at pH 7) a current density of 10 mA/cm$^2$ at 1.85 V versus RHE (Figure 3b, black squares) whereas a 50 mV higher potential was required for $IrO_2$-$RuO_2$ to show the same current density (Figure 3b, gray triangles). Therefore Tafel lines that can be assigned to both samples move (toward higher potentials) towards each other in the lower overpotential region (<500 mV) and move (toward higher potentials) apart from each other in the higher overpotential region (>500 mV) respectively (Figure 3b). Interestingly, the Tafel line belonging to sample Ni42-300 was substantially stiffer in the lower overpotential (slope 198.16 mV dec$^{-1}$) region than the one assigned to $IrO_2$-$RuO_2$ (slope 345.1 mV dec$^{-1}$). A substantial horizontal shift (~160 mV) of the Tafel line of sample Ni42-300 (black squares) compared to the corresponding Tafel line of untreated steel Ni42 (gray circles) towards lower potentials (Figure 3b) proves the meaningful enhancement of the OER relevant electrocatalytic properties under neutral conditions upon the applied surface oxidation. This horizontal shift between Tafel lines assigned to the Ni42 sample and the Ni42-300 sample can be explained by a change in the chemical nature of the surface during surface oxidation. An increase in the active area caused by the surface oxidation, which could be one possible reason for higher currents at specific potentials, would lead to a vertical shift of the CV curves towards higher current densities. However we checked the surface area of sample Ni42 and sample Ni42-300 by performing multiple point nitrogen gas adsorption BET measurements (Figures S3a/S3b) as well as AFM investigations (Figures S4). As expected there are no substantial differences regarding surface area (Ni42: 0.354 m$^2$/g; Ni42-300: 0.37 m$^2$/g), or the adsorption/desorption plots of both samples (Figure S3a/S3b), and sample Ni42-300 (roughness 29.538 nm) proved to be even smoother than sample Ni42 (71.521 nm) (Figures S3).

Single Tafel slopes could be obtained throughout the potential region (IrO$_2$-RuO$_2$: 345.10 mV dec$^{-1}$; Ni42: 150.88 mV dec$^{-1}$; Ni42-300: 198.16 mV dec$^{-1}$). Tafel slopes of OER electrocatalysts determined in pH 7 solutions have rarely been published, but very recently graphene-Co$_3$O$_4$ nanocomposites and ZrS$_3$ nanosheets were reported as prospective OER electrocatalysts for water splitting under neutral and alkaline conditions. For these materials Tafel slopes of 98 mV dec$^{-1}$ respectively 102 mV dec$^{-1}$ in 0.1 M phosphate buffer solution [31, 89] were determined. We investigated oxidized standard carbon manganese steel S235 as a potential OER electrocatalyst in pH 7 regime and found at lower potentials a Tafel-slope of 172.4 mV dec$^{-1}$, which is comparable to the slopes reported here and those slopes reported in the literature [30, 31, 33].

*Origin of the layer formation-The catalytic active species*. We recently showed that the Ni enrichment in the outer sphere of AISI 304 steel during the electroactivation (300 min, 1.77 A/cm$^2$, 7.2 M NaOH) very likely occurs due to a dissolution mechanism rather than upon an electro migration process [45]. However electromigration, as a potential driving force for the layer formation, could not be excluded with absolute certainty, and may at least reinforce the observed changes of the composition of the surface of AISI 304 steel during anodization [45]. This role of electromigration applies substantially more for the data we present in the current study. Whereas AISI 304 steel revealed a considerable mass loss (~ 7.1 mg) via electro-activation [45], the average mass loss detected after electro-oxidation of Ni42 steel amounted to 0.092 mg (Table S1), which is over 70 times less than in the case of activated AISI 304 steel[45]. Due to the relatively high amount of electrolyte (200 mL) the concentration of Fe, Ni and Mn was below the detection limit of our ICP OES device, and in addition did not leave sufficient material for an analysis via Atomic Absorption Spectroscopy (AAS) as performed in our recent study [45]. Neither Ni nor Fe ions could be detected upon an analysis "*by hand*" (Supporting information) of the electrolyte used for electro-activation. XPS experiments (results shown in the main text; Figure 4) were carried out with our samples after polarization experiments: cyclic voltammetry plus chronopotentiometry measurements carried out for t< 4000s). Additional results from XPS studies are shown in the supporting information (Figure S4): The spectra shown were recorded after 40000 s of OER upon sample Ni42-300 at 4 mA/cm$^2$ in pH 7 solution (200OER), after 40000 s of HER at 10 mA/cm$^2$ in pH 13 solution (200HER) respectively. A further extension of the polarization tests up to duration of 450000 s did not influence the XPS findings summarized in Figure S4. The composition of the electrode periphery derived from the cationic distribution on the surface of sample Ni42-300 (80.8% Ni, 18.55% Fe, 0.57% C) confirmed our expectation that a Ni containing compound is the predominant cationic species of the surface oxidized Ni 42 alloy (Table S2). In direct comparison with untreated alloy (sample Ni42) the Ni content was substantially increased from 26.6 at. % to 80.8 at. % whereas the iron content was found to be substantially decreased (Table S2). Due to the lack of mass loss (Table S1) during electro-activation, and the absence of hints that would support the hypothesis that dissolution at least of some ingredients of the steel takes place (no coloration of the electrolyte, no Ni, Fe ions in the electrolyte, no layer deposited on the counter electrode) we conclude that electromigration of Ni to the surface is the most likely origin for the Ni enrichment and Fe depletion in the outer sphere of the activated steel obtained during anodization. Electromigration, which means a mass transport caused by a momentum transfer e- → M+ at high current densities [61] is supposed to be responsible for the growth of thin layers and was obtained by Medway *et al.* for a film growth on Ni metal during electrochemical cycling [62].

In one of our previous reports dedicated to electrocatalytically supported oxygen formation upon oxidized steel surfaces we determined, on the basis of XPS studies, γ–NiO(OH) as the potentially catalytic active species, or at least the dominating source for catalytic active species on the periphery of the treated steel for OER in alkaline medium (0.1 M KOH) [45]. Bediako *et al.* investigated OER at pH 9.2 upon Ni containing films that have been generated via electrodeposition starting from Ni$^{2+}$containing borate solutions [63]. Detailed, XANES and EXAFS based structure-activity correlations indicated that it is very likely that γ–NiO(OH) is the catalytic active OER species in this Ni containing film under more neutral conditions (pH 9.2). These data challenged the long-held notion that the beta phase of NiO(OH) is a more efficient catalyst. Therefore based on our findings for OER at pH 13 [45] and the findings of other groups for OER at pH 9.2, we expected to also detect γ–NiO(OH) as the catalytic active OER species on the surface of sample Ni42-300 during water splitting in 0.1 M phosphate buffer solution (pH 7).

Figures 4a, b display high resolution Ni 2p and Fe 2p spectra of Ar etched, as well as untreated Ni42 and Ni42-300 samples.

The 2p$_{3/2}$ positions of some reference compounds [64, 65, 66, 69, 70] are indicated by gray vertical bars. The Ni 2p spectrum of the untreated sample Ni42 (Figure 4a) contains both metallic and oxidized Ni. The Ar etched Ni42 showed predominantly unoxidized Ni, whereas the Ar etched Ni42-300 exhibited metallic (Ni) fractions plus oxidized species that we determined as γ–NiO(OH). Ar ion etching obviously reduced +3 oxidized Ni because Ni$^0$ is missed in the surface of untreated Ni42-300 similar to data reported by Leinen *et al.* [67]. The significantly enhanced signal to noise ratio and higher resolved satellite structure of the spectrum for the non-etched sample Ni42-300 (compared to that of untreated sample Ni42) is due to the higher Ni concentration on the surface of the sample and can clearly be seen in Figure 4a. The presence of metallic Ni and Ni(II), species such as NiO or Ni(OH)$_2$, on the surface of sample Ni42-300 can be rather excluded. To our experience very harsh oxidative conditions of the anodization procedure (300 min, 2 A/cm$^2$ current density) completely converts Ni$^0$ in Ni based steels into Ni(III) species [45]. γ–NiO(OH) was found to be the dominating Ni species on the surface of anodized Ni42 steel and this confirms our expectation that Ni species are responsible for the electrocatalytic activity regarding oxygen formation upon the surface in aqueous media. This agrees very well with a recent *operando* X-ray absorption spectroscopy (XAS) study in which (Ni, Fe)OOH catalysts were characterized during OER operating conditions [68]. From the Fe 2p spectra (Figure 4b) we find a small metallic fraction for all samples. The iron of sample Ni42 (untreated) may be oxidized to Fe$_2$O$_3$ (Fe$^{3+}$) at the surface and is likely to be caused by long exposure to air during storage. Due to Ar ion etching, Fe$^{3+}$ on the surface of sample Ni42 was reduced to Fe$^{2+}$ and Fe$^0$ as can be determined from the Fe 2p spectrum of Ni42 recorded after the etching procedure (Figure 4b). There is no significant difference between the outcome of the Fe 2p-XPS analysis of sample Ni42-300, non-etched and Ar etched Ni42-300 (Figure 4b). The presence of Fe$^{2+}$ can be excluded given the binding energies of the Fe2p$_{3/2}$ core level spectra. With respect to the Fe 2p$_{3/2}$ binding energies of sample Ni42-300 the FeOOH species of iron dominates on the surfaces compared to the peak positions found for the reference compounds [65, 69, 70] indicated by vertical bars in Figure 4b. The results of the XPS studies carried out after non steady state- or short time OER chronopotentiometry experiments (Figure 4a/b) are in agreement with the findings obtained after 40000 s of OER at 4 mA/cm$^2$ in pH 7 medium (see Figure S5a,b; black curves), i.e. the electrocatalytic active species remained the same.

The roles that Fe species (cationic distribution of Ni42-300:80.8% Ni, 18.55% Fe) may play in enhancing the oxygen evolution from water upon Ni-based electrocatalysts is to date not fully understood. In our recent study dedicated to activated AISI 304 steel we were able to prove a strong dependence of the electrocatalytic OER properties at pH 13 on the Ni:Fe relation, and found optimal OER performance in case of a cationic distribution of 67% Ni and 33% Fe [45]. Several groups including Corrigan *et al.* [71] and Trotochaud *et al.* [85] already discussed the effects of Fe incorporation on Ni/Fe oxyhydroxide thin films. In addition to full-experimental approaches computational based results have also been exploited [72]. Generally, in case of thick layers a substantial portion of the applied potential during OER polarization experiments will drop across the catalyst film to drive the transport of electrons through it when the layer is not sufficiently conductive. As stated in these reports Fe embedded in NiO(OH) films will reduce the resistivity and thus for thick films (> 1 µm) Fe impurities is a prerequisite for low overpotentials determined in electrochemically initiated OER. On the basis of these findings we assume that in our case an "incorporation" of 18.5% Fe into the Ni-oxide based layer (Table S2) is required to ensure a low conductivity of the relatively thick layer (~4.7 µm, See Figure S2) and will lead to the low overpotentials obtained by us. An activity enhancement of the catalytically active Ni ions in our samples upon partial-charge transfer between them and Fe ions resulting in the formation of $Ni^{3+/4+}$ with increased oxidizing power as shown by various groups for the binary Ni-Fe-oxides [73] could also play a role. For instance a charge transfer was confirmed between ultrathin NiO(OH) layers and gold substrate by Yeo and Bell [78]. The electron withdrawing effect of highly oxidized Fe ($Fe^{3+}$) should be similar to noble metals and is likely to be the origin of an increase in OER activity of NiO(OH) doped with $Fe^{3+}$ as obtained by Trotochaud *et al.* who used NiO(OH) films 40-60 nm in thickness [85]. As can be taken from Figure 4 b Fe in oxidation state +3 indeed dominates also on the surface of sample Ni42-300.

In addition it is long been a discussion about the transition between for instance $\gamma$–NiO(OH) and $\beta$–NiO(OH) due to unintentional- or intentional Fe incorporation and its effect on the electrocatalytic OER efficiency [74, 75]. Redox phase transitions between Fe(III)/Ni(III)/Ni(IV) oxyhydroxides in Fe-Ni OER electrocatalysts at oxidative potentials have been obtained very recently upon exploiting ambient-pressure X-ray photoelectron spectroscopy (APXPS) [76].

**2.2 OER properties in alkaline medium.** Given the composition of the surface of our oxidized steel Ni42 (Table S2), and based on our studies on OER under alkaline conditions upon modified Ni-Fe alloys [45] we expected strong electrocatalytic OER performance of sample Ni42-300 also under alkaline conditions. We evaluated the OER properties of sample Ni42-300 in 0.1 M KOH and 1 M KOH on the basis of non-steady state (Figures 5a, b) and steady state (Figures 5c,d,e) electrochemical characteristics. The significant improvement of the OER behavior upon surface modifications of Ni42 steel can be seen in the cyclic voltammetry studies (Figure 5a/5b) in which sample Ni42-300 (Figure 5b, black curve) exhibited a much higher current density to voltage ratio than sample Ni42 over the entire voltage range (Figure 5a). The current density in 0.1 M KOH at 1.7 V *vs.* RHE amounted to around 31 mA/cm$^2$ (Figure 5b, black curve) for sample Ni42-300 whilst for untreated Ni42 the current density was about half (16 mA/cm$^2$, (Figure 5a) at the same potential. The CV curve of untreated Ni 42 recorded at pH 13 (Figure 5a) is similar to the one recorded at pH 7 (Figure 1b), again neither the characteristic Ni(II)-Ni(III) oxidation wave was obtained, nor any peak in the cathodic part of the CV could be assigned to typical Ni(III)-Ni(II) transition. In contrast, the CV curve of sample Ni42-300 showed at a potential of ~ 1.4 V *vs.* RHE, the expected Ni(II)-Ni(III) oxidation wave (Figure 5b, black curve), and in the cathodic part at a potential of 1.3 V. *vs.* RHE also showed the corresponding Ni(III)-Ni(II) reduction wave (Figure 5b, black curve). An overpotential of 254 mV was detected from 2000 s of chronopotentiometry using sample Ni42-300 at 10 mA/cm$^2$ current density (Figure 5c, black curve) in 0.1 M KOH. This presents an OER performance similar to the one determined for OER at pH 13 on AISI 304 steel, electro-oxidized in a comparable way (269.2 mV overpotential at 10 mA/cm$^2$)[45] and is significant superior to the OER properties of both, chlorinated AISI 304 steel (260 mV at 1.5 mA/cm$^2$)[23] or chlorinated S235 steel (347 mV overpotential at 2.0 mA/cm$^2$) [33] at pH 13. At this pH value sample Ni42-300 exhibited stronger voltage current ratios within OER polarization experiments than $IrO_2$-$RuO_2$ ($\eta$= 351 mV at j=10 mA/cm$^2$; Figure 5c, green curve). The OER performance of Ni42-300 at pH 13 is on a similar level to that of non-noble metal based state of the art catalysts recently developed by other groups including $Co_3O_4$ nanoparticles [77, 78], $Co_2P$ nanoparticles [79], -$NiCo_2O_4$ –graphene hybrids [80]; $Ni_3S_2$ nano arrays supported by Ni metal [81]; $CuFe(MoO_4)_3$[82] ;$Pr_{0.5}Ba_{0.5}CoO_3$ [83]; $NiCo_2O_4$ aerogels [84]; $Ni_xFe_y(OH)_2$[85] and $Ni_xFe_yO_z$ nanoparticles supported on glassy carbon, [44] all of which are significantly more expensive and complex to produce. The OER performance of our catalyst Ni42-300 did not suffer upon extension of the operating time up to 450 ks (Figure S1, blue curve): $\eta$ amounted to 251 mV through 450 ks of OER in pH 13 regime at 10 mA/cm$^2$. As expected, substantial stronger voltage-current behavior in polarization tests was achieved when OER was performed in higher concentrated KOH instead of 0.1 M KOH (pH 13). This can be seen in the graphical representation Figure 5d that comprises a CV (blue curve) and a chronopotentiometric plot derived from pH 14 measurements (magenta curve). A current density as high as 60 mA/cm$^2$ was reached at 1.525 V *vs.* RHE corresponding to 297 mV overpotential (Figure 5d, blue curve). This represents a six times higher current density compared to that achieved within the CV plot recorded in 0.1 M KOH at the same potential (Figure 5b, blue curve). The corresponding 2000 s chronopotentiometric scan exhibited averaged $\eta=215$ mV at 10 mA/cm$^2$ (Figure 5d, magenta colored curve). To check the performance and stability of the catalyst under more realistic industrial conditions OER was performed in 7 M KOH at 70 °C (Figure S6/S7). The potential required to ensure 10 mA/cm$^2$ current density was found to be even reduced throughout the 450000 s of chronopotentiometry ($\eta=196$ mV; Figure S6). Under repeated cycling of the potential the degradation of electrocatalysts is usually found to be increased. However a stronger current-voltage behavior can be derived from the CV recorded after 1000 cycles when compared to the one determined at the beginning of the experiment (Figure S7), thus confirming our expectation that the OER performance of sample Ni42-300 will become even better during usage in OER experiments. After 1000 cycles a current density of 130 mA/cm$^2$ was reached at 1.5 V *vs.* RHE (dashed curve in Figure S7) whereas the first scan showed 115 mA/cm$^2$ at 1.5 V *vs.* RHE (black curve in Figure S7). This increase of the OER activity of an electrocatalyst upon strong usage is initially unexpected. However the group of Chatenet made similar obtainments for 316 steel based electrodes in lithium-air batteries [26]. The Tafel line created on the basis of chronopotentiometry measurements performed with sample Ni42-300 (Figure 5e, blue squares) was significantly negatively shifted compared to the one derived from sample $IrO_2$-$RuO_2$ in 0.1 M KOH (Figure 5e, green squares), underpinning the overall superior OER performance of surface oxidized Ni42 alloy. The difference in overpotential required for a defined current density between both samples increases substantially with increasing current density. The corresponding Tafel lines move apart from each other towards the higher potential regions (Figure 5e). The corresponding slopes of the lines also differ substantially: the Tafel line of sample Ni42-300 exhibited a slope of 71.6 mV dec$^{-1}$ whereas the slope of the Tafel line of $IrO_2$-$RuO_2$ amounted to 101.1 mV dec$^{-1}$ (Figure 5e). Doyle and Lyons found for OER on passive oxide covered iron electrodes Tafel slopes of 40 mV dec$^{-1}$ [86] at lower potentials.

As can be taken from previous reports, Tafel lines of surface activated steel samples showed in the lower potential region slopes of 49 mV dec$^{-1}$ (AISI 304)[45], 58.5 mV dec$^{-1}$ (S235)[33] whereas untreated steel showed significant higher slopes of 66 mV dec$^{-1}$ (AISI 304)[45] and 102 mV dec$^{-1}$ (S235)[33] with one exception: 30 mV dec$^{-1}$ (AISI 316 in 1 M KOH) [87]. In addition the value for the Tafel slope of sample Ni42-300 in 0.1 M KOH (71.6 mV dec$^{-1}$) is similar to recently reported Tafel slopes of NiFe alloy [88], iron electrodes [89], carbon supported NiO nanoparticles [44] and for "fresh" Ni metal surfaces. [89] Recently developed bi-functional electrocatalysts exhibited slopes between 52 and 89 mV dec$^{-1}$ for the OER at pH 14 [79, 110, 111, 112].

IrO$_2$-RuO$_2$ showed at pH 13 Dual Tafel behavior (within the investigated potential region) with lower slopes at lower overpotential regions (101.1 mV dec$^{-1}$) and higher slopes at higher overpotential regions (Figure 5e). The value determined for sample Ni42-300 (71.6 mV dec$^{-1}$ ≙ 2.1 x RT/F) agrees well with the value that was found to be characteristic of an O$_2$ evolution mechanism involving a reversible one-electron transfer (2.3 x RT/F) [90]. In summary, supported by the electrochemical measurements, our surface oxidized Ni-42 steel proved to be an OER electrocatalyst that is, highly active in neutral as well as in alkaline conditions (pH 7, 13, 14 and 14.6).

**2.3 OER properties in acidic medium.** Ni based alloys are famous for their excellent OER properties in alkaline regimes [91]. Generally very few metal oxides can survive under oxidative potentials in acidic regimes [92] and metal oxide based catalysts have very rarely been checked for catalytic activity under both, alkaline and acidic conditions. Again Rutile type IrO$_2$ and RuO$_2$ seem to be the material of choice when good water splitting properties are desired at low (<< 7) and high (>>7) pH values [93, 94, 92, 95, 96]. In terms of the overall OER activity and stability under acidic catalysis conditions IrO$_2$ is considered as the best compromise [96]. In order to evaluate the anodic water-splitting properties in acidic regimes we performed cyclic voltammetry studies on sample Ni42-300 in 0.05 M H$_2$SO$_4$ from which an overpotential of ~360 mV at 1 mA/cm$^2$ current density can be revealed (Figure S8). An overpotential of ~ 200 mV for the OER on mesoporous templated IrO$_2$ surfaces in 0.5 M H$_2$SO$_4$ was reported by Strasser *et al.*[96]. However we did not detect reasonable steady state behavior for the OER at pH 1 initiated on the surface of sample Ni42-300. The sample was found to be unstable in pH 1 medium at positive potentials for longer time and exhibited even at very low potentials (~0.2 V *versus* RHE) significant current densities. This finding agrees well with previous studies where McCrory *et al.* evaluated Ni-oxide, Co-oxide and mixed oxides consisting of different non-noble transition metals as well as IrO$_2$ as prospective OER catalysts in 1 M H$_2$SO$_4$. Every system tested apart from IrOx was unstable under oxidative conditions in acidic solutions [92]. The pronounced solubility of metal oxides that were formed on the surface of Ni42 metal during the electro-activation process is, in our view, a reasonable potential explanation for the poor stability of sample Ni42-300 under oxidative conditions in diluted sulfuric acid.

**2.4 HER properties in acidic/alkaline medium**. OER is considered as the more challenging water-splitting electrode reaction and it is considered worth to be a focus of optimization efforts [97, 98, 99]. In part certainly also motivated by the fact that hydrogen, with a price of 8 €/kg, is the more valuable water-splitting product, improving HER also has a high research status. The electrochemically initiated hydrogen evolution on metal surface is well established. More than one hundred years ago Tafel *et al.* investigated the electrode kinetics [100, 101], with more detailed reaction mechanisms suggested in follow up studies, and hydrogen evolution activities quantified by Trasatti *et al.* [102, 103] respectively Miles et al [104].

It should be noted for all HER measurements presented in this work cathodic currents were applied, i.e. even if not explicitly stated, all values of current densities carry a negative sign. Even untreated Ni42 exhibited measurable HER activity (5.7 mA/cm$^2$ at -400 mV *versus* RHE in 0.05 M $H_2SO_4$; Figure 6a, red curve). Nevertheless the voltage-current characteristics in the voltage range one typically has to sweep over for HER (-450-0 mV *vs.* RHE) were improved substantially upon our chosen surface modification (11 mA/cm$^2$ at 400 mV *versus* RHE in 0.05 M $H_2SO_4$; Figure 6a, blue curve). The performance difference between the electrochemical HER properties of surface modified and non-modified Ni42 examined on the basis of non-steady state electrochemical measurements was even more pronounced at pH 13 (Figure 6b). Sample Ni42 was found to be relatively inactive towards electrocatalytically initiated HER in alkaline media as seen in the weak current to voltage behavior in the corresponding cyclic voltammogram (Figure 6b, red curve). Whilst moving from positive towards negative potentials the increment of the current density was found to be moderate. At the limit of our measuring interval of -0.4 V *versus* RHE the current density amounted to less than 2 mA/cm$^2$ (Figure 6b, red curve). In stark contrast, the CV curve of sample Ni42-300 was found to be much stiffer-showing a significant increase of the current density towards higher overpotentials (Figure 6b, blue curve) and almost 15 mA/cm$^2$ current density was reached at -0.4 V *versus* RHE (Figure 6b, blue curve). To ensure sufficient comparability with earlier published results the electrochemical behavior was also examined in more concentrated electrolytes. Significantly higher current densities were achieved for the HER of sample Ni42-300 when CVs were recorded in tenfold higher concentrated solutions (0.5 M $H_2SO_4$, 1 M KOH; Figure 6c). HER initiated in 0.5 M $H_2SO_4$ on sample Ni42-300 resulted at -0.4 V *versus* RHE in 60 mA/cm$^2$ current density and required for 10 mA/cm$^2$ current density $\eta$=220 mV (Figure 6c, blue curve). The sample was found to be less active for HER in 1 M KOH leading to ~ 30 mA/cm$^2$ current density at $\eta$=400 mV, 10 mA/cm$^2$ current density at $\eta$=325 mV respectively (Figure 6c, black curve). Chronopotentiometry- and Tafel plots for HER in $H_2SO_4$ (0.05 M and 0.5 M) and in KOH (0.1 M plus 1 M) allowed to evaluate the stability of sample Ni42-300, whilst electrocatalytically initiated $H_2$ formation under acidic (pH 1 and 0) and alkaline conditions (pH 13 and 14) and are shown in Figures 7 a-c/9a, b. A reliable assessment of the HER efficiency can only be extracted from the outcome of these steady-state measurements. Sample Ni42-300 proved to be a stable HER electrocatalyst for tens of hours of chronopotentiometry at 1 or 10 mA/cm$^2$ at pH 1 and also at pH 13 (Figure 7c), at pH 0 and 14 respectively (data not shown). Even after more than 100 hours operating time the HER performance in the pH 1 regime was similar to that at the start level (Figure S9). Repeated cycling of the potential of Ni42-300 between +25 and -200 mV *vs.* RHE in pH 1 medium increased the HER performance of Ni42-300 showing the very good stability of the sample towards HER in acidic regime (Figure S10). This result is by far not self-evident. Ni-Mo alloys, famous for their low overpotentials for HER in alkaline media at industrially relevant current densities [47, 48, 49, 50, 51] were indeed found to be highly active towards HER in acidic solutions, too, but Ni and Ni-Mo alloys, generally all known remaining earth-abundant solid catalysts, exhibited rather poor catalytic stability upon hydrogen formation under acidic conditions [52]. Based on the data obtained from the non-steady state measurements (Figures 6 a-c) we expected to determine (under steady state conditions in the lower overpotential region) a somewhat better HER performance, i.e. lower overpotentials for the HER in acidic solution. Indeed sample Ni42-300 exhibited $\eta$=240.1 mV at 1 mA/cm$^2$ at pH 13 (Figure 7a, black curve) whereas at pH 1 the overpotential was substantially lower ($\eta$=128.6 mV at 1 mA/cm$^2$ current density Figure 7b; black curve). At 10 mA/cm$^2$ current density the overpotential for the HER amounted to 283.6 mV (pH 13) and 316.5 mV (pH 1) respectively (Figures 7a/b, blue curves). The values for the HER overpotential of sample Ni42-300 derived from long term (50000 s) chronopotentiometric measurements carried out at 10 mA/cm$^2$ in 0.1 M KOH ($\eta$=333 mV at pH 13; Figure 7c, blue curve) and 0.05 M H$_2$SO$_4$ ($\eta$=321 mV at pH 1; Figure 7c, red curve) did not significantly differ from those values derived from 1000 s chronopotentiometry plots (Figures 7a/b, blue curves). These electrochemical HER experiments demonstrate the very good stability of sample Ni42-300 under HER catalysis conditions, i.e. under reductive potentials in alkaline and acidic regimes.

We therefore expected a quantitative charge to hydrogen formation, and this was confirmed by determining the Faradaic efficiency for the HER upon Ni42-300 at pH 13 because- the efficiency amounted to 101.8% as seen in Figure 8. Typically, the charge to hydrogen conversion rate of up to date bifunctional electrocatalysts amounts to ~ 100% in alkaline solutions [109, 110, 111, 112] with one exception: Co$_2$P nanowires exhibited a HER faradaic efficiency of ~80% in 1 M KOH [79].

Log j *versus* $\eta$ plots derived from catalyzed HER performed at pH 1 and 13 (Figure 9a) and at pH 0 and 14 (Figure 9b) revealed single Tafel behavior within a reasonable voltage range (-400-0 mV *versus* RHE). Tafel slopes determined in less diluted electrolytes differed not much from the ones recorded in more dilutes electrolytes (Figure 9b, grey hexagons; 0.5 M H$_2$SO$_4$: 117.9 mV dec$^-$ [1]; Figure 9b, black squares, 1 M KOH: 117.47 mV dec$^{-1}$). In diluted media Tafel slopes of sample Ni42-300 amounted to 80.47 mV dec$^{-1}$ (0.05 M $H_2SO_4$), 124.44 mV dec$^{-1}$ (0.1 m KOH) respectively (Figure 9a, gray hexagons and black squares). Tafel slopes for the HER in acidic regime have been determined by various groups. First investigated by Trasatti *et al.* Tafel slopes for catalyzed HER via $RuO_2$ in 1 M $HClO_4$ amounted to 60 mV dec$^{-1}$ [105]. Very low Tafel slopes have also been reported for $IrO_2$ (30 mV dec$^{-1}$), $RuO_2$ and mixed Ru-Rh-oxides (40 mV dec$^{-1}$) for HER-electrocatalysis carried out in 2.5 M $H_2SO_4$ [106] as well as in 1 M $H_2SO_4$ (40 mV dec$^{-1}$ for $RuO_2$) [107]. Nanostructured non-noble transition metal based HER electrocatalysts exhibited at reasonable overpotentials ($\eta$= 150-200 mV) slightly higher Tafel slopes ($MoS_2$: 94 mV dec$^{-1}$ in 0.5 M $H_2SO_4$ [108]; $Ni_2P$: 81 mV dec$^{-1}$ in 0.5 M $H_2SO_4$ [53]. Bi-functional electrocatalysts investigated very recently exhibited for the HER in 1 M KOH slope values between 84 and 141 mV dec$^{-1}$ [110, 111, 112], as such close to the one of Ni42-300 shown for the HER at pH 14 (117.47 mV dec$^{-1}$).

Given the strong current to voltage ratio in more concentrated acidic- or alkaline solutions extracted from cyclic voltammogram`s (Figure 6c) we expected to determine from steady state data significant lower overpotentials for the HER at pH 0 and pH 14 when compared to the values derived from chronopotentiometry performed at pH 1 and pH 13 (Figures 7a,b, c). As a matter of fact an overpotential as low as 189 mV at 10 mA/cm$^2$ current density was determined for electro-catalyzed HER on Ni42-300 surface in 0.5 M $H_2SO_4$ (pH 0, Figure 9b gray hexagons), and 299 mV overpotential at 10 mA/cm$^2$ current density for HER upon sample Ni42-300 in 1 M KOH (pH 14) respectively (Figure 9b, black squares). In addition, the suitability of Ni42-300 as hydrogen evolving electrode in industrial electrolyzers was proven under harsh conditions (7 M KOH, 70 °C) (Figures S11/S12) exhibiting very good stability. Upon 450 ks of chronopotentiometry at 10 mA/cm$^2$ current density the overpotential decreased from 330 mV to 230 mV (Figure S11). A slightly better performance of sample Ni42-300 during heavy usage is also reflected in the comparison of the cyclic voltammograms gained after one- and after a thousand cycles (Figure S12). Other groups found that incorporation of nonmetals like Sulfur, Nitrogen, Carbon into Ni based alloys appeared to be an encouraging strategy to increase the efficiency and the stability of materials towards HER in acidic solutions [52]. For instance Schaak *et al.* reported on the outstanding electrocatalytic activity (100 mV overpotential at 10 mA/cm$^2$ current density) and stability of nanostructured $Ni_2P$ towards HER in 0.5 M $H_2SO_4$ [53]. However a catalyst capable of catalyzing both HER and OER in different or even the same media with appropriate performance is still difficult to find. Martindale *et al.* reported on bi-functional iron only electrodes for full water splitting at pH 13 at a bias of ~ 2V [109]. Significant lower cell voltages of 1.65 V, 1.68 V and 1.63 V were required for overall water splitting upon CoSe[110] film, NiCo2S4 nanowires[111] and NiSe films [112] in 1 M KOH as shown by T. Liu *et al.*, D Liu *et al.* and C. Tang *et al.* However, the scarcity of selenium can certainly be considered as an obstacle for practical usage. Very recently Stern *et al.* [113] and Antonietti and Shalom et. al. [54] published data on $Ni_2P$ nanoparticles (NPs) respectively $Ni_5P_4$ NPs as Janus catalyst in alkaline regime. Besides the already mentioned HER properties under acidic conditions [53] relatively high OER activity (290 mV overpotential) and high HER activity (~ 200 mV overpotential) combined with sufficient stability under catalysis conditions could be attested to $Ni_2P$ NPs in 1 M KOH at 10 mA/cm$^2$ as well [113]. However Schaak *et al.* achieved a contradicting result and reported on quickly degraded $Ni_2P$ to Ni when used as a HER electrocatalyst in 1 M KOH [53]. A comparable HER/OER activity and very good stability under catalysis conditions in 1 M KOH was reported for $Ni_5P_4$ NPs (HER: $\eta$= 150 mV at 10 mA/cm$^2$ ; OER: $\eta$= 330 mV at 10 mA/cm$^2$ ) [54]. Table S4 gives an idea of the overall electrocatalytic OER- and HER properties of recently developed bifunctional electrocatalysts.

Although the OER performance of $Ni_2P$- as well as $Ni_5P_4$ NPs in 1 M KOH is not on benchmark level the combination of outstanding HER and adequate OER properties makes both compounds to attractive catalysts. However, especially in direct comparison with these very recently published data our catalyst constitutes in our opinion a highly promising alternative given the unrivaled low production costs, the very reasonable HER performance, and the stability at pH 14 ($\eta$=299 mV at 10 mA/cm$^2$) and also at pH 0 ($\eta$=189 mV at 10 mA/cm$^2$), especially as this is achieved in combination with the outstanding OER activity and stability at pH 13 ($\eta$=254 mV at 10 mA/cm$^2$) respectively at pH 14 ($\eta$=215 mV at 10 mA/cm$^2$).

**The catalytic active species whilst HER upon Ni42-300.** Figure S5 represents the outcome of our XPS study performed with the sample Ni42-300 after long term OER- (200OER) and HER (200HER) chronopotentiometry. Whereas the curve form of the Fe 2p core level spectrum of 200OER and 200 HER do not significantly differ (Figure S5b), the corresponding Ni 2p core level spectra show indeed slight differences (Figure S5a). The Ni 2p core level spectrum of 200HER was found to be shifted towards lower binding energy and Ni(II) species like NiO instead of $\gamma$-NiOOH is likely the predominant catalytic active species during HER upon the surface of Ni42-300 (Figure S5a). A further increase of the HER operating time up to 450000 s did not influence the XPS results.

## 3. Conclusion

Improving the conversion of water into its cleavage products $H_2$ and $O_2$ photo- or electrocatalytically realized using renewable energy sources is a key challenge to overcome the limited availability of fossil fuels for future applications. This conversion route is especially promising if a catalyst is suitable to catalyze both, OER and HER, and ideally with both electrode exhibiting half reactions in one single electrolyte. An identical catalyst for OER and HER catalytically initiated in the same medium, which is classically termed *Full Water Splitting Catalyst* would pave the way for the implementation of an electrolyzer module with bipolar cell configuration thus substantially simplifying the design of a water-splitting device. The studies reported here evaluate the suitability of AISI Ni42 steel, surface oxidized upon an electrochemical approach, as a water-splitting OER and HER electrocatalyst in alkaline, neutral and acidic solutions. We found that anodization under harsh conditions (2000 mA/cm$^2$ current density; duration: 300 min, 7.2 M KOH) converted relatively less active untreated Ni 42 steel into an outstanding OER and reasonable HER electrocatalyst. The layer, firmly attached to the alloy matrix formed during this kind of activation basically consists of γ-NiO(OH), $Fe_2O_3$ besides FeO(OH). The most significant results of the work described can be highlighted as:

(1) Ni 42 steel, electro-activated for 300 min under harsh conditions (sample Ni42-300) exhibited unusual high catalytic activity combined with very good long term (for the entire 450000 s tested) stability when used as an OER electrocatalyst in pH 7 phosphate buffer solution given an average overpotential of 491 mV at 4 mA/cm$^2$ in pH 7 corrected 0.1 M $KH_2PO_4$ /$K_2HPO_4$ through 40000 s chronopotentiometry plot. The Ni-Fe-oxide layer, around 5 μm in thickness, generated on the alloy during electro-oxidation sufficiently protected the matrix against further- (inner) oxidation highlighted by an outstanding Faradaic efficiency of sample Ni42-300 of 99.4 % (4000 s of chronopotentiometry at 2 mA/cm$^2$) that makes the OER properties even superior to those of $IrO_2$-$RuO_2$.

(2) Sample Ni42-300 also proved to be highly active and stable (up to the 450000 s maximum tested) towards OER electrocatalysis under alkaline conditions (under steady state and non-steady state conditions) at pH 13 (0.1 M KOH) and 14 (1 M KOH) and exhibited significantly better electrocatalytic efficiency than $IrO_2$-$RuO_2$. The catalytic OER key figures of Ni42-300 were η=251 mV at 10 mA/cm$^2$ in 0.1 M KOH, and η=215 mV at 10 mA/cm$^2$ in 1 M KOH. In addition the catalyst exhibited very good activity and outstanding stability (η=196 mV at 10 mA/cm$^2$ through 450000s chronopotentiometry plot) for OER performed

(3) Sample Ni42-300 was found to be well suitable as an HER electrocatalyst with regards to activity and stability in aqueous solutions over a wide pH range spanning from pH 0-14.6. In diluted acidic- and alkaline media (0.05 M $H_2SO_4$ and 0.1 M KOH) the overpotential for HER amounted to around 268.4 mV (pH 1) and 333 mV (pH 13) to guarantee a stable current density of 10 mA/cm$^2$. Long term chronopotentiometry measurements carried out at 10 mA/cm$^2$ current density for 450000 s of HER (pH 1), for 50000 s of HER (pH 13) respectively, exhibited sufficient stability under catalysis conditions as shown by the non-existent increment of the potential through the 450000 s plot (pH 1), 50000 s plot (pH 13) respectively. The current to voltage ratio, within the cyclic voltammetric- and chronopotentiometric studies was much stronger when those were carried out in 1 m KOH and 0.5 M $H_2SO_4$ instead of 0.1 M KOH and 0.05 M $H_2SO_4$ respectively ($\eta$= 299 mV at 10 mA/cm$^2$ at pH 14, $\eta$= 189 mV at 10 mA/cm$^2$ at pH 0). No signs of degradation of the catalyst were obtained when HER polarization tests carried out in 7 M KOH at 70 °C for long operating times up to the maximum 450000 s tested.

In summary surface oxidized Ni42 alloy combines good HER performance and stability at pH 14.6 ($\eta$=275 mV at 10 mA/cm$^2$), pH 14 ($\eta$=299 mV at 10 mA/cm$^2$), pH 13 ($\eta$=333 mV at 10 mA/cm$^2$), pH 1 ($\eta$=268.4 mV at 10 mA/cm$^2$) and pH 0 ($\eta$=189 mV at 10 mA/cm$^2$) with outstanding OER activity and stability at pH 7 ($\eta$=491 mV at 4 mA/cm$^2$), pH 13 ($\eta$=251 mV at 10 mA/cm$^2$), pH 14 ($\eta$=215 mV at 10 mA/cm$^2$ and at pH 14.6 ($\eta$=196 mV at 10 mA/cm$^2$). Especially when compared with very recently achieved Janus type characteristics of $Ni_2P$ [113] and $Ni_5P_4$ [54] our catalyst therefore constitutes a highly promising alternative Full Water Splitting catalyst. The unrivaled cost efficiency of a water-splitting device consisting of modified Ni42 electrodes is not only based on the inexpensive Ni42 alloy but also on the potentially possible, cost effective bipolar cell configuration.

## 4. Experimental Section

**Preparation of the Ni 42-300 samples-Electro oxidation of stainless steel at constant potential**

Samples with a total geometry of 45x10x1,5 mm were constructed from 1,5 mm thick AISI Ni 42 steel. Pre-treatment: Prior to each surface modification the surface of the metal was cleaned intensively with ethanol and polished with grit 600 SiC sanding paper. Afterwards the surface was rinsed intensively with deionized water and dried under air for 100 min. The weight was determined using a precise balance (Sartorius 1712, 0.01 mg accuracy) prior to electro-activation.

For the electro-oxidation a two-electrode set-up was used consisting of the steel sample as WE, and a platinum wire electrode (4x5 cm) used as CE. The WE (anode) was immersed exactly 2.1 cm deep (around 4.5 cm$^2$ geometric area), and the CE (cathode) was completely immersed into the electrolyte.

The electrolyte was prepared as follows: In a 330 mL glass beaker, 57.6 g (1.44 mol) of NaOH (VWR, Darmstadt, Germany) was dissolved under stirring and under cooling in 195 g deionized water. The solution was allowed to cool down to 23°C before usage. The anodization was performed under stirring (450 r/min) using a magnetic stirrer and a stirring bar (21 mm in length, 6 mm in diameter). The distance between WE and CE was adjusted to 6 mm. A power source (Electra Automatic, Vierssen, Germany) EA-PSI 8360-15T which allows to deliver a constant voltage even at strongly changing current loads was used for the electrochemical oxidation. The procedure was carried out in current controlled mode. The current was set to 8.7 A according to 2000 mA/cm$^2$ current density. The voltage varied during the electro-activation. At the beginning of the experiment it amounted to around 6.7 V but was reduced within the duration of 300 min to around 4.5 V. These data proved to be reliably reproducible for all 10 replicates. If however for some reasons the decrease of the voltage is more abrupt, i.e. the voltage drops down to 4,5 V earlier, the activation procedure should be stopped when a voltage of 4.5 V is reached. The temperature of the electrolyte increased within the first 30 min and reached a value of 323 K. After every hour approximately 2.5 mL of fresh 6 M NaOH was added to the electrolysis vessel in order to compensate the loss occurred due to evaporation. After 300 min of electro-activation the CE and the WE were taken out of electrolyte and rinsed intensively with tap water for 15 min and afterwards with deionized water for a further 10 min. The used NaOH electrolyte was transferred quantitatively into a plastic bottle and was stored for further analysis. The counter electrode was immersed into 30 mL of 2 M HCl for 12 hours. The acidic solution was stored for further analysis via ICP OES. Prior to the electrochemical characterization the samples were dried under air at ambient temperature and the weight was determined upon a precise balance as described above. The sample preparation was repeated ten times, i.e. in total 11 samples of Ni 42-300 have been prepared this way.

## Supporting Information

Supporting Information is available from the Wiley Online Library or from the author

**Acknowledgements**: D.M.C. was supported by the NSERC CGS-Alexander Graham Bell scholarship and P.Z. acknowledges the NSERC Discovery Grant for funding. A special thank you is given to Andrew George (Dalhousie University, Department of Physics) for his technical assistance during XPS experiments. H. S., M. S. and M.S. were supported by the European Research Council (ERC-CoG-2014; project 646742 INCANA).

Author contribution statement

HS had the idea to perform the experiments the manuscript is based on. He planned and performed the electrochemical measurements and all sample preparations. HS wrote the manuscript as well as the sup. Information. DMC, PZ, KK and JW planned, performed and evaluated the XPS measurements. K. M.-B. and JS planned and performed the BET measurements. JDH helped in the evaluation of some experiments. HS, SS, UK, MS, SD and LW performed and evaluated the AFM as well as the FIB/SEM experiments. All authors have read the manuscript.

**Conflict of Interest Disclosure**: The authors declare no competing financial interest.

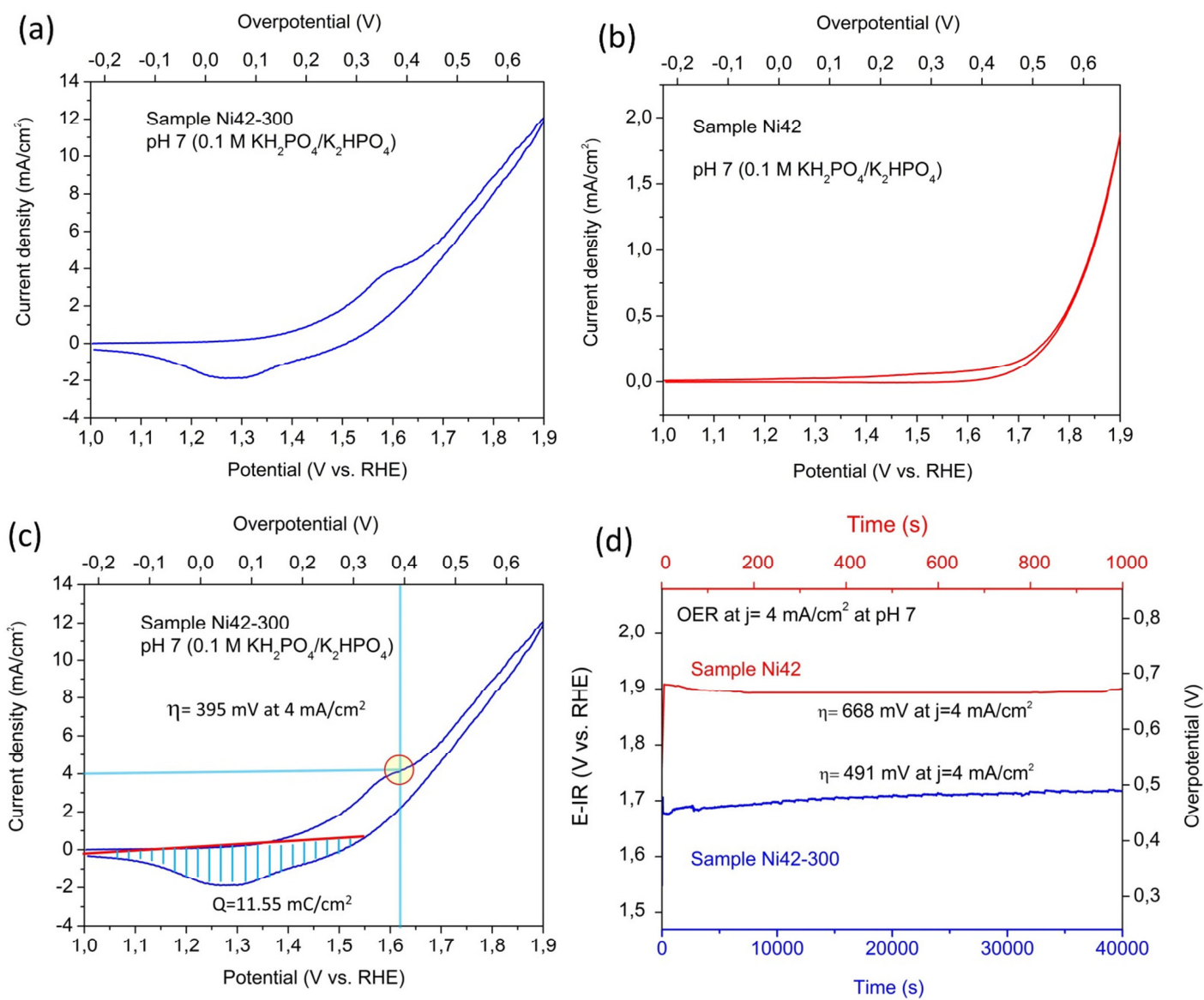

Figure 1. Overview of the electrochemical OER properties of non-treated Ni 42 alloy (sample Ni42) and surface oxidized Ni42 (sample Ni42-300) at pH 7. Cyclic voltammetric plots are based on 20 mV/s scan rate and 2 mV step size. Electrode area of all samples: 2 cm$^2$. (a) Cyclic voltammogram of sample Ni42-300 in pH 7 corrected 0.1 M $KH_2PO_4/K_2HPO_4$. (b) Cyclic voltammogram of sample Ni42 in pH 7 corrected 0.1 M $KH_2PO_4/K_2HPO_4$. (c) Cyclic voltammogram of sample Ni42-300 in pH 7 corrected 0.1 M $KH_2PO_4/K_2HPO_4$. Determination of the charge capacity Q was achieved by integrating the cathodic voltammetric sweep between 1.55 V *vs.* RHE and 1 V *vs.* RHE. Q amounted to 11.55 mC cm$^{-2}$. (d) Long term chronopotentiometric measurement of samples Ni42 and Ni42-300 performed in pH 7 corrected 0.1 M $KH_2PO_4/K_2HPO_4$ at 4 mA/cm$^2$ current density. Average overpotential for the OER through 40000 s plot: 491 mV (Ni42-300); 668 mV (Ni42).

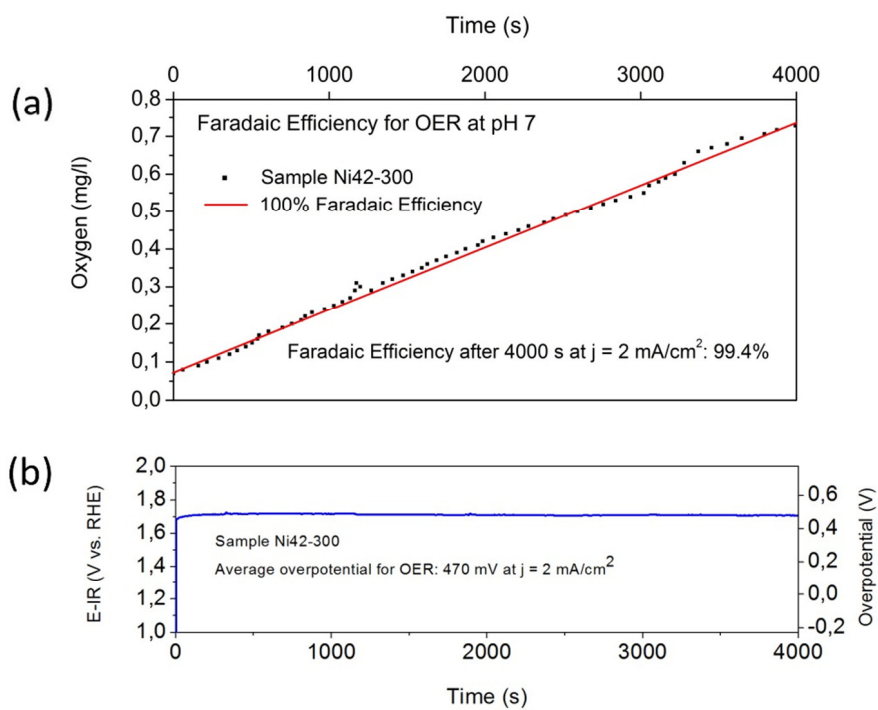

Figure 2. Faradaic Efficiency of the OER upon sample Ni42-300 at pH 7 whilst chronopotentiometric measurement performed for 4000 s at 2 mA/cm$^2$ current density. Electrode area of sample Ni42-300 was 2 cm$^2$. **(a)** Correlation of oxygen evolution upon sample Ni42-300 in 0.1 M $K_2HPO_4/KH_2PO_4$ (dotted curve) determined with a optical dissolved oxygen (OD) sensor using the so-called fluorescence quenching method with the charge passed through the electrode system (grey line corresponds to 100% Faradaic efficiency). Amount of the electrolyte: 2.0 l; Start value of dissolved oxygen: 0.07 mg/l (t= 0 s); End value of dissolved oxygen (t = 4000 s): 0.729 mg/l (nominal value (100%):0.7333 mg/l). Faradaic efficiency of the OER after 4000 s runtime: 99.4 %. **(b)** Corresponding chronopotentiometry plot. Current density: 2 mA/cm$^2$; Average Overpotential for the OER through the 4000 s plot: 470 mV.

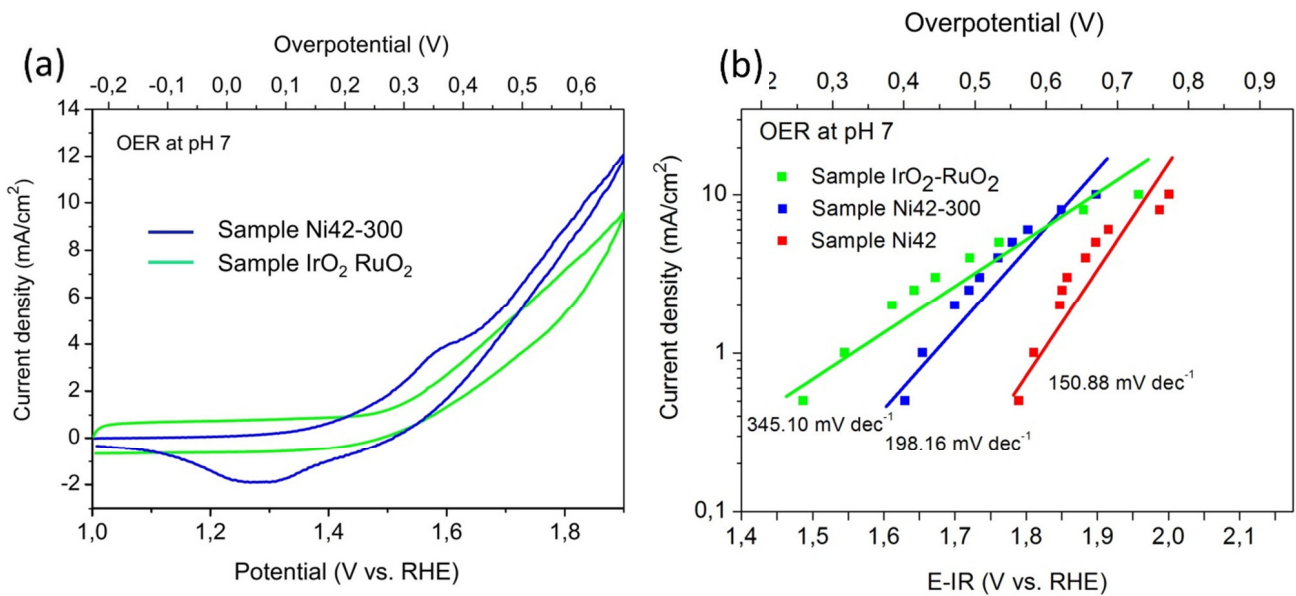

Figure 3: Comparison of the electrocatalytic non-steady state and steady state OER properties of samples Ni42-300, Ni42 and $IrO_2$-$RuO_2$ (PVD sputtered IrO2-RuO2 layers, ≥ 10 μm in thickness on Titanium) in pH 7 corrected $K_2HPO_4$/$KH_2PO_4$ solution. Electrode area of all samples: 2 cm$^2$. (a) Cyclic voltammograms of samples Ni42-300 (black curve) and $IrO_2$-$RuO_2$ (gray curve). Measurements were carried out with 20 mV/s scan rate and 2 mV step size. (b) Tafel plots of samples Ni42-300, Ni42 and $IrO_2$-$RuO_2$ based on 200 second chronopotentiometry scans at current densities of 0.5, 1, 2, 2.5, 3, 4, 5, 6, 8 and 10 mA/cm$^2$ current density.

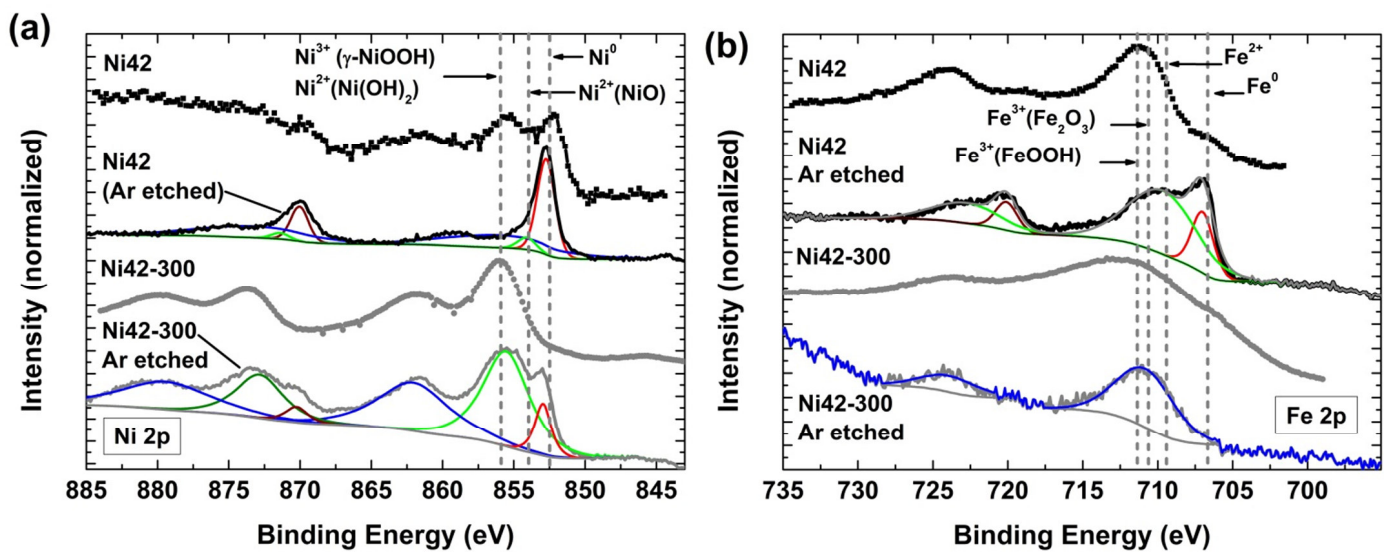

Figure 4. High resolution XPS spectra of samples Ni42 and Ni42-300 untreated as well as Ar etched. (a) Ni 2p 3/2 and Ni 2p /1/2 core level spectra. Fitting results for Ni42 (Ar etched):Ni 2p 3/2, position 852.735 eV; Ni 2p½, position 870.035 eV; Ni2p 3/2, position 854.099 eV; Ni 2p ½, position 871.399 eV; Ni 2p 3/2, position 855.988 eV; Ni 2p ½, position 873.288 eV. Fitting results for Ni42-300 (Ar etched): Ni 2p 3/2, position 852.949 eV; Ni 2p ½, position 870.249 eV; Ni2p 3/2, position 855.535 eV; Ni 2p ½, position 872.835 eV; Ni 2p 3/2, position 862.034 eV; Ni 2p ½, position 879.413 eV. (b) Fe 2p 3/2 and Fe 2p /1/2 core level spectra. Fitting results for Ni42 (Ar etched):Fe 2p 3/2, position 706.997 eV; Fe 2p 3/2, position 709.67 eV; Fe 2p ½, position 720.097 eV; Fe 2p ½, position 722.77 eV. Fitting results for Ni42-300 (Ar etched): Fe 2p 3/2, position 710.813 eV; Fe 2p 1/2, position 723.913 eV. Binding energies of reference compounds [64, 65, 66, 67, 69] are indicated by vertical lines as visual aid.

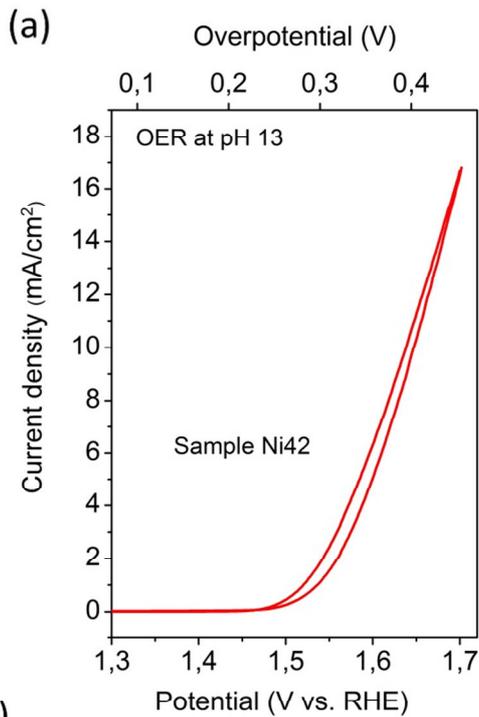
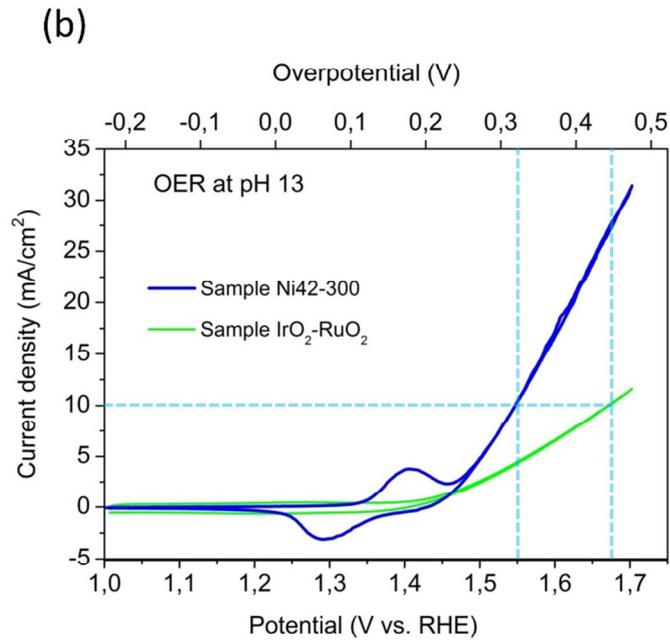
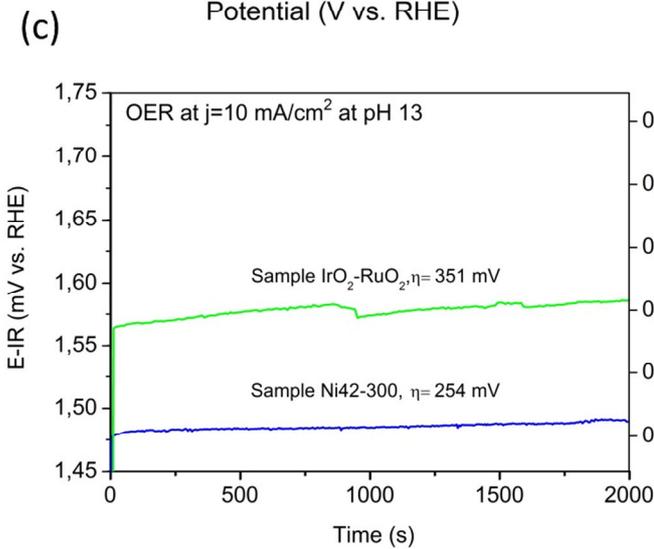
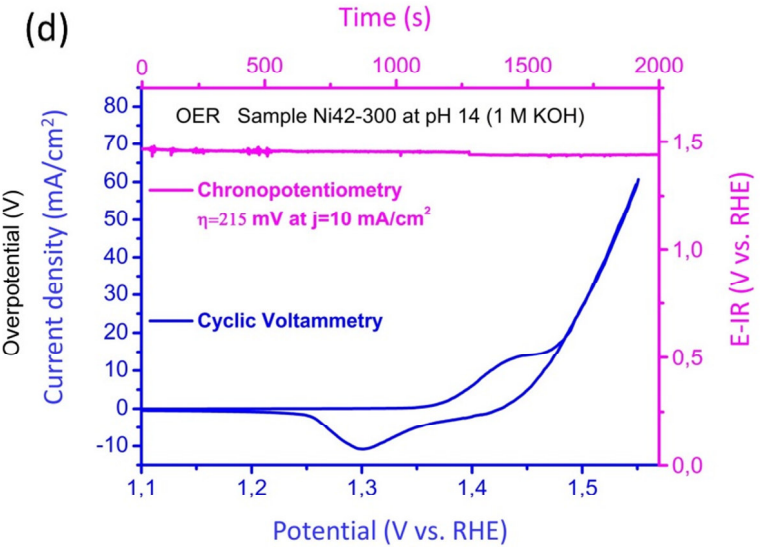
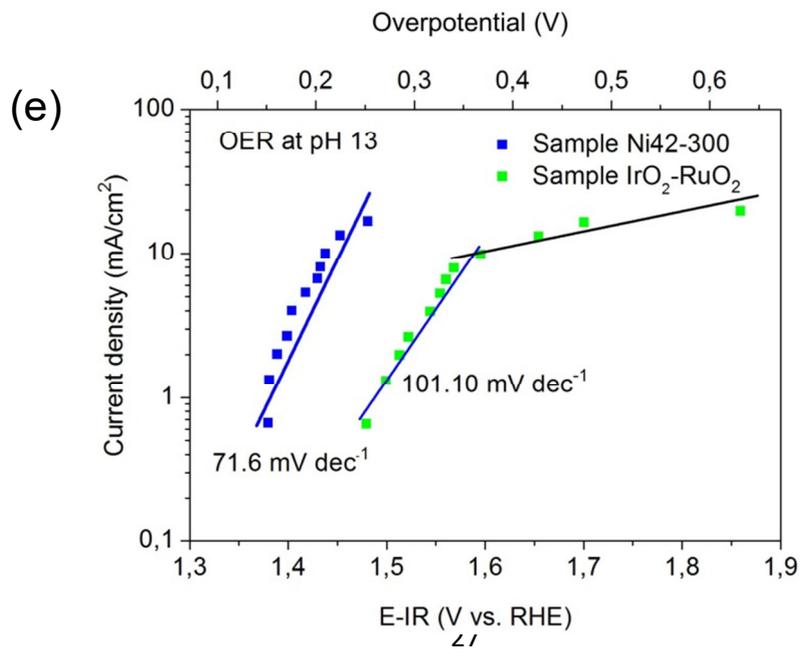

Figure 5. Steady-state and non-steady state OER properties of sample Ni42, Ni42-300 and $IrO_2$-$RuO_2$ at pH 13 and 14. Electrode area of all samples: 2 $cm^2$. Cyclic voltammetric plots are based on 20 mV/s scan rate and 2 mV step size. (a) Cyclic voltammogram of sample Ni42 recorded in 0.1 M KOH. (b) Cyclic voltammogram`s of samples Ni42-300 (black curve) and $IrO_2$-$RuO_2$ (gray curve) in 0.1 M KOH. (c) Chronopotentiometry measurements performed with sample Ni42-300 (black curve) and $IrO_2$-$RuO_2$ (gray curve) in 0.1 M KOH at 10 mA/$cm^2$ current density. Average overpotential for the OER through 2000 s scan: 351 mV ($IrO_2$-$RuO_2$), 254 mV (Ni42-300). (d) Cyclic voltammogram of sample Ni42-300 recorded in 1 M KOH (black curve) and 2000 s chronopotentiometric plot of sample Ni42-300 determined in 1 M KOH at 10 mA/$cm^2$ (gray curve); Average overpotential for the OER through 2000 s plot: 215 mV. (e) Tafel plots of samples Ni42-300 and $IrO_2$-$RuO_2$ based on 200 second chronopotentiometry scans in 0.1 M KOH at current densities of 0.66, 1.33, 2, 2.67, 4, 5.33, 6.67, 8, 10, 13.33, 16.67 and 20 mA/$cm^2$.

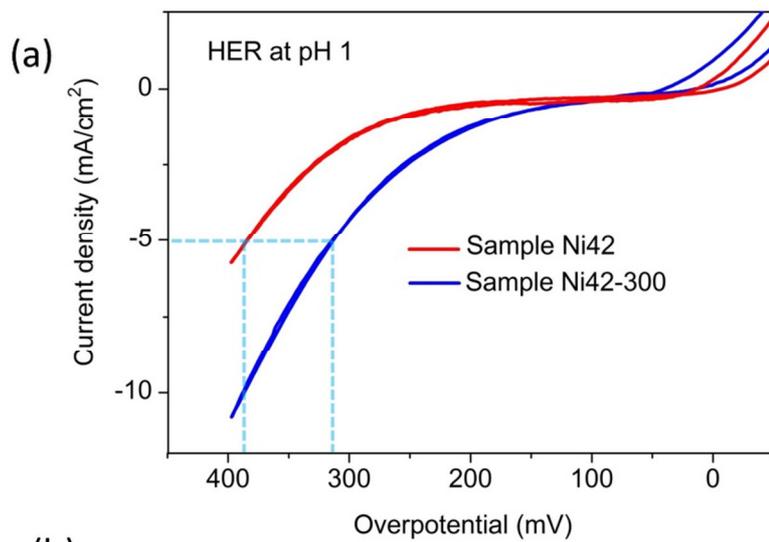

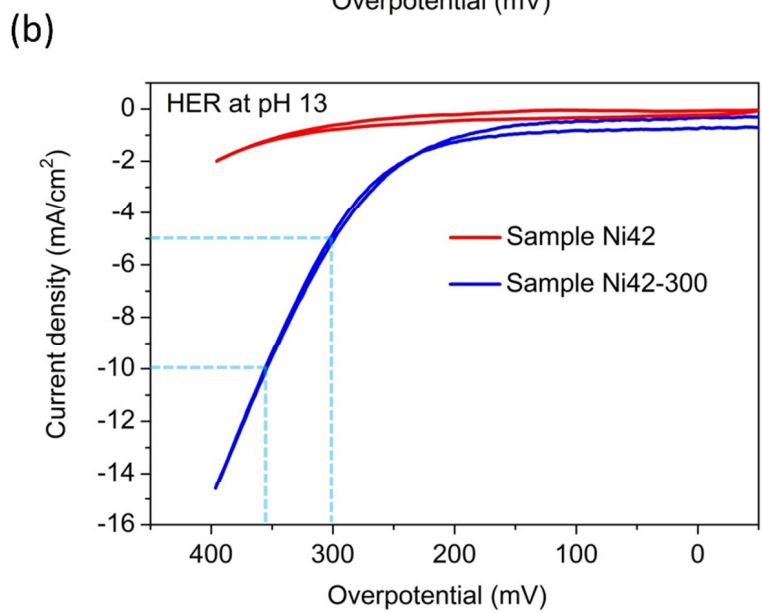

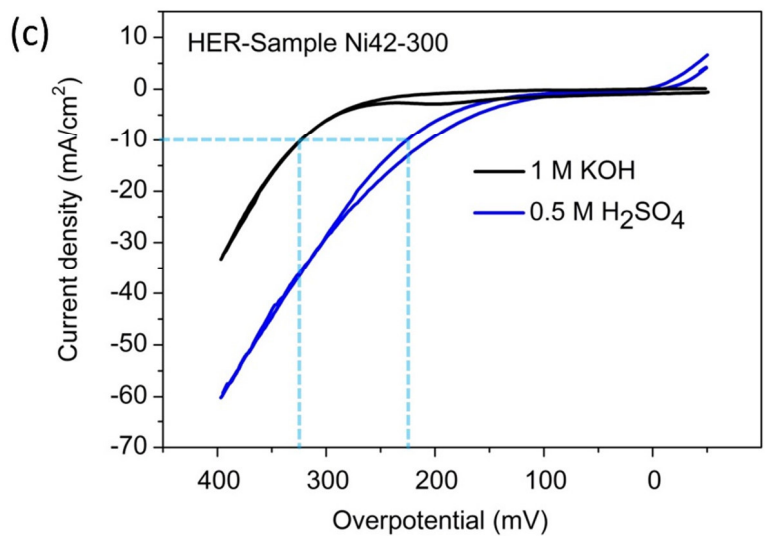

Figure 6. Comparison of the non-steady state HER properties of samples Ni42 and Ni42-300 at pH 1 and pH 13, of sample Ni42-300 at pH 0 and pH 14 respectively. Settings: 5 mV/s scan rate, 2 mV step size. Electrode area of all samples: 2 cm$^2$. (a) Cyclic voltammogram`s of samples Ni42 (gray curve) and Ni42-300 (black curve) recorded in 0.05 M $H_2SO_4$. (b) Cyclic voltammogram`s of samples Ni42 (gray curve) and Ni42-300 (black curve) recorded in 0.1 M KOH. (c) Cyclic voltammogram`s of sample Ni42-300 recorded in 0.5 M $H_2SO_4$ (gray curve) and in 1 M KOH (black curve).

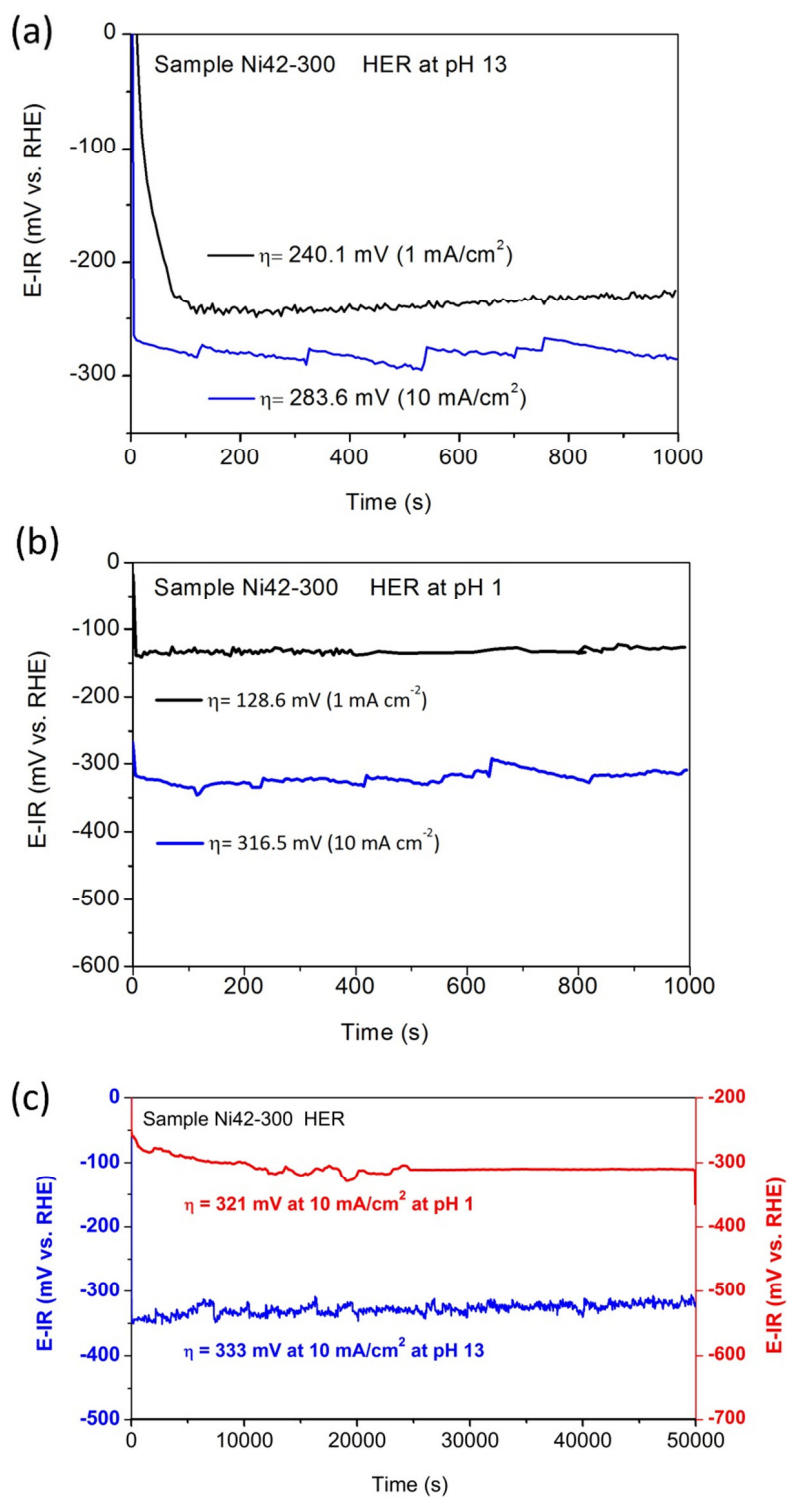

Figure 7. Steady state HER properties of sample Ni42-300 at pH 1 and 13. Electrode area: 2 cm$^2$. Cathodic currents were applied, i.e. even if not explicitly stated, all values of HER current densities carry a negative sign (a) Chronopotentiometry measurements performed in 0.1 M KOH at 1 mA/cm$^2$ (black curve), at 10 mA/cm$^2$ (gray curve) respectively. Average overpotential for the HER through 1000 s scan: 240.1 mV (1 mA/cm$^2$), 283.6 mV (10 mA/cm$^2$) respectively. (b) Chronopotentiometry measurements performed in 0.05 M H$_2$SO$_4$ at 1 mA/cm$^2$ (black curve), at 10 mA/cm$^2$ (gray curve) respectively. Average overpotential for the HER through 1000 s scan: 128.6 mV (1 mA/cm$^2$), 316.5 mV (10 mA/cm$^2$) respectively. (c) Long term chronopotentiometric measurement of sample Ni42-300 performed at 10 mA/cm$^2$ current density in 0.05 M H$_2$SO$_4$ (gray curve), in 0.1 M KOH (black curve) respectively. Average overpotential for the HER through 50000 s plot: 321 mV (0.05 M H$_2$SO$_4$), 333 mV (0.1 M KOH) respectively.

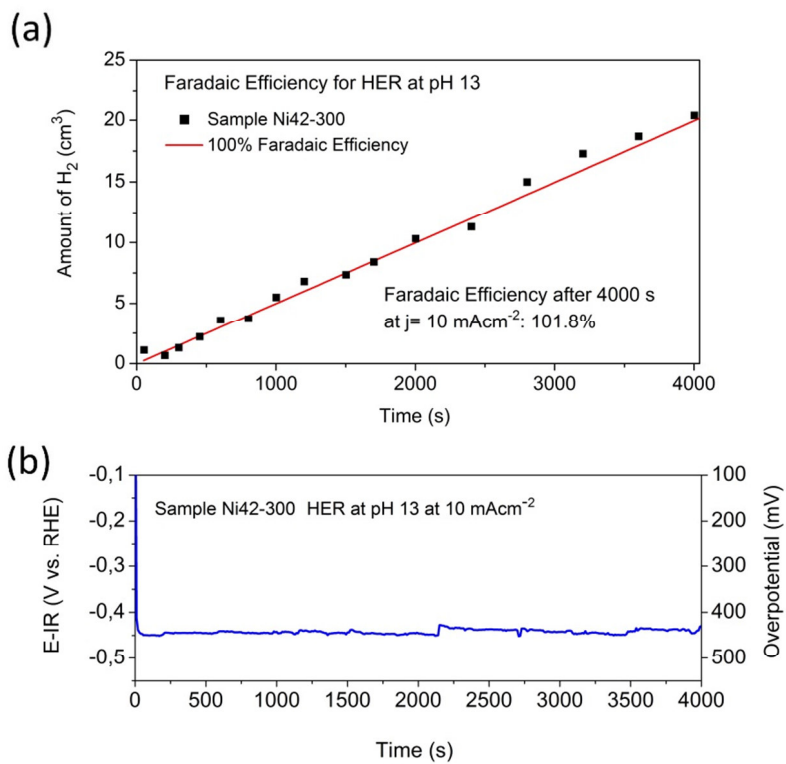

Figure 8. Faradaic Efficiency of the HER upon sample Ni42-300 at pH 7 during chronopotentiometric measurement performed for 4000 s at 10 mA/cm$^2$ current density. Electrode area of sample Ni42-300 was 2 cm$^2$; Amount of the electrolyte: 1.7 l **(a)** Correlation of oxygen evolution upon sample Ni42-300 in 0.1 M KOH (dotted curve) with the charge passed through the electrode system; gray line corresponds to 100% Faradaic efficiency with a line equation: y=4.9863 x 10$^{-3}$x with y=amount of hydrogen (cm$^3$); x=time (s). End value of oxygen amount (t = 4000 s): 20.3 mL (nominal value (100%):19.945 mL). Faradaic efficiency of the OER after 4000 s runtime: 101.8 %. **(b)** Corresponding chronopotentiometry plot. Current density: 10 mA/cm$^2$; Average Overpotential for the HER through the 4000 s plot: 428 mV.

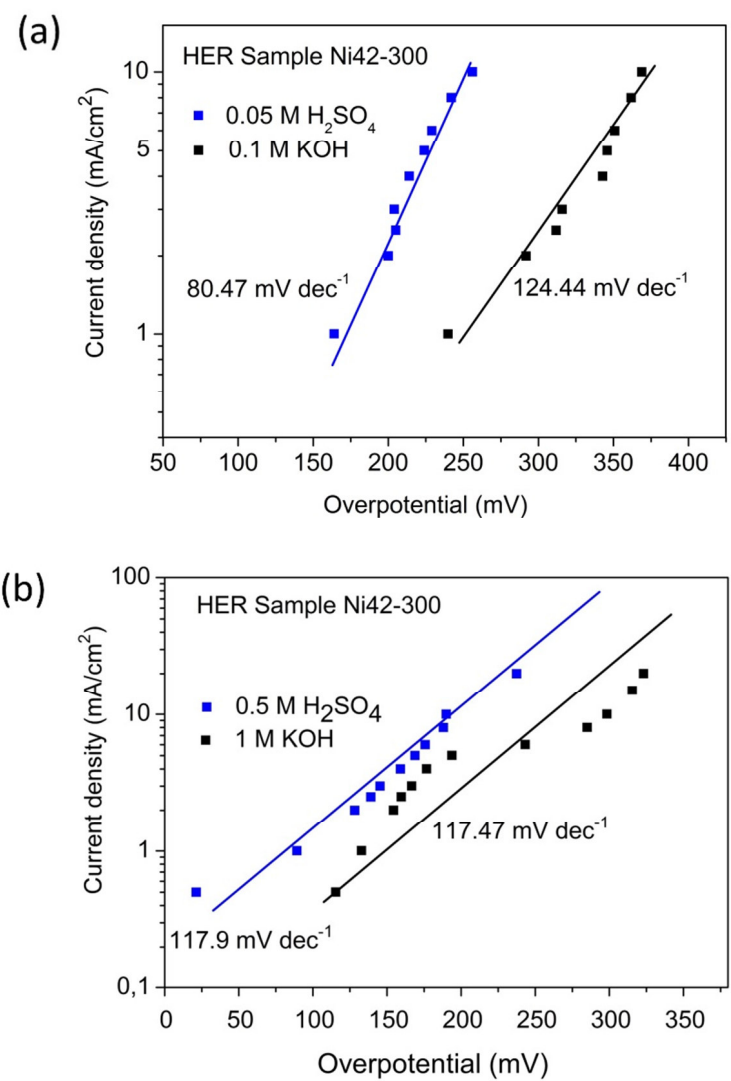

Figure 9 (a) Tafel plots of sample Ni42-300 based on 200 second chronopotentiometry scans in 0.05 M $H_2SO_4$ (gray line) and 0.1 M KOH (black line) at current densities of 1, 2, 2.5, 3, 4, 5, 6, 8 and 10 mA/cm$^2$. (b) Tafel plots of sample Ni42-300 based on 200 second chronopotentiometry scans in 0.5 M $H_2SO_4$ (gray line) and 1 M KOH (black line) at current densities of 0.5, 1, 2, 2.5, 3, 4, 5, 6, 8, 10 and 20 mA/cm$^2$.

# Supporting Information

# Electro Oxidation of Ni42 Steel: A highly Active Bifunctional Electrocatalyst

Helmut Schäfer*[a], Daniel M. Chevrier [b], Peng Zhang [b], Johannes Stangl [c], Klaus Müller-Buschbaum [c], Jörg D. Hardege[d], Karsten Kuepper[e, f], Joachim Wollschläger [e, f], Ulrich Krupp[g], Simon Dühnen [a], Martin Steinhart [a], Lorenz Walder [a], Shamaila Sadaf [a], and Mercedes Schmidt[a]

[a] *Institute of Chemistry of New Materials and Center of Physics and Chemistry of New Materials, Universität Osnabrück, Barbarastrasse 7, 49076 Osnabrück, Germany*

[b] *Department of Chemistry, Dalhousie University, Halifax, Nova Scotia, Canada B3H 4J3*

[c] *University of Würzburg, Institute of Inorganic Chemistry Julius-Maximilians-Universität Würzburg Am Hubland, D-97074 Würzburg, Germany*

[d] *School of Biological, Biomedical and Environmental Sciences, University of Hull, Cottingham Road, Hull, HU6 7RX, United Kingdom*

[e] *Department of Physics, Universität Osnabrück, Barbaraßtraße 7, 49069 Osnabrück, Germany*

[f] *Center of Physics and Chemistry of New Materials, Universität Osnabrück, Barbaraßtraße 7, 49069 Osnabrück, Germany*

[g] *Institute of Materials Design and Structural Integrity University of Applied Sciences Osnabrück, Albrechtstraße 30, 49076 Osnabrück, Germany*

**Electrochemical Measurements**

A three-electrode set-up was used for all electrochemical measurements. The working electrode (WE) with a total geometry of 45x10x1,5 mm was constructed from 1,5 mm thick AISI Ni 42 steel or AISI Ni42-300 sample, prepared as described before, on which an apparent surface area of 2 cm$^2$ was defined by an insulating tape (Kapton tape). AISI Ni42 was purchased from Schmiedetechnik Faulenbach, Wiehl, Germany. The IrO$_2$-RuO$_2$ sample (10 micrometer layer deposited on titanium) with a total geometry of 100x100x1.5 mm was purchased from Baoji Changli Special Metal Co, Baoji, China. An electrode area of 2 cm$^2$ was defined on the plate by Kapton tape. To avoid additional contact resistance the plate was electrically connected via a screw. A platinum wire electrode (4x5 cm geometric area) was employed as the CE, a reversible hydrogen reference electrode (RHE, HydroFlex, Gaskatel Gesellschaft für Gassysteme durch Katalyse und Elektrochemie mbH. D-34127 Kassel, Germany) was utilized as the reference standard, therefore all voltages are quoted against this reference electrode (RE). For all measurements the RE was placed between the working and the CE. The measurements were performed in a pH 7 corrected 0.1 M KH$_2$PO$_4$/K$_2$HPO$_4$, solution, 0.1 M-, 1 M- and 7 M KOH (VWR,

Darmstadt, Germany), 0.05 M $H_2SO_4$, 0.5 M $H_2SO_4$ (VWR, Darmstadt, Germany) solution respectively. Measurements were performed at room temperature (295.15 K) with the exception of the measurements carried out in 7 M KOH (343.15 K). The pH 7 corrected 0.1 M $KH_2PO_4$/$K_2HPO_4$ solution was prepared as follows: Aqueous solutions of 0.1 M $K_2HPO_4$ and $KH_2PO_4$ (VWR, Darmstadt, Germany) were mixed until the resulting solution reached a pH value of 7.0. The distance between the WE and the RE was adjusted to 1 mm and the distance between the RE and the CE was adjusted to 4-5 mm. All electrochemical data were recorded digitally using a Potentiostat Interface 1000 from Gamry Instruments (Warminster, PA 18974, USA), which was interfaced to a personal computer. All non-steady state electrochemical measurements were carried out without any correction of the voltage drop. Regarding the steady state electrochemical measurements: we corrected Ohmic losses within the chronopotentiometry plots manually by subtracting the Ohmic voltage drop from the measured potential on the basis of electrolyte resistances reported in literature[114] (Table S3). The resistances reported for 10 mm RE-WE distance were corrected accordingly taking into consideration the different electrode geometry, i.e. 1 mm distance between WE and RE. IR-corrected potentials are denoted as E-IR. **Cyclic Voltammograms (CV)** were recorded in 90 mL of electrolyte (pH 7 corrected 0.1 M $KH_2PO_4$/$K_2HPO_4$, 0.1 M KOH, 1 M KOH, 7 M KOH, 0.05 M $H_2SO_4$, 0.5 M $H_2SO_4$) in a 100 mL glass beaker under stirring (450 r/min) using a magnetic stirrer (21 mm stirring bar). The scan rate was set to 20 mV/s and the step size was 2 mV for the OER related measurements and the scan rate was set to 5 mV/s and the step size was 2 mV for the HER related measurements. The potential was cyclically varied between 1 and 1.9 V *vs.* RHE for OER measurements at pH 7, between 1 and 1.7 V *vs.* RHE for OER measurements at pH 13, between 1 and 1.57 V *vs.* RHE for OER measurements at pH 14 and between 1 and 1.50 V *vs.* RHE for OER measurements at pH 14.6. The potential was cyclically varied between +50 and -400 mV *vs.* RHE for all HER related measurements with the exception of measurements according to Figures S9 and S11 (+25 mV>E>-200 mV).

**Chronopotentiometry scans** were conducted depending on the nature of the electrolyte at a constant current density of 1, 4 respectively of 10 mA/cm$^2$ in 90 mL of electrolyte for measuring periods < 2000 s, in 800 mL of electrolyte for measuring periods ≥ 40000 s, in 1200 mL of electrolyte for a measuring period of 450000 s in a 100 mL, 1000 mL respectively 1500 mL glass beaker. The scans were recorded under stirring (450 r/min) using a magnetic stirrer (25 mm stirring bar) for measuring periods < 2000 s, using a magnetic stirrer (40 mm stirring bar) for measuring periods ≥ 40000 s respectively.

**Tafel plots**

Average voltage values for the Tafel plots were derived from 200 second chronopotentiometry scans at current densities of 0.5, 1, 2, 2.5, 3, 4, 5, 6, 8 and 10 mA/cm$^2$ for measurements carried out at pH 1, 7 and 13, at current densities of 0.66, 1.33, 2, 2.67, 4, 5.33, 6.67, 8, 10, 13.33, 16.67 and 20 mA/cm$^2$ for measurements carried out at pH 0 and pH 14, respectively. The arrangement of RE, WE and CE (taken for recording the chronopontentiometry plots) was as mentioned above (See paragraph *Electrochemical measurements*).

**Determination of Faradaic efficiency for OER** (Figure 2a, b) was carried out in close accordance with the procedure described in Schäfer *et al.*, *Energy Envirnon. Sci.*, **2015**, DOI:10.1039/C5EE01601K (ref. 45). Faradaic efficiency of OER was calculated by determining the dependence of the oxygen concentration in the electrolyte during the time of chronopotentiometry at constant current of 2 mA/cm$^2$ in 0.1 M K$_2$HPO$_4$/KH$_2$PO$_4$ solution under stirring. The distance between RE and WE was adjusted to 1 mm and the distance between RE and CE was adjusted to 4-5 mm. The volume of electrolyte was 2000 mL. The working compartment was completely sealed with glass stoppers before starting the chronopotentiometry at 0.07 mg O$_2$/l. The results can be taken from **Figure** 2a. The red line in **Figure** 2a corresponds to 100% Faradaic efficiency with a line equation: y=0.0001658220884x + 0.07 with y=Dissolved oxygen (mg/l); x=time (s).

**Determination of Faradaic efficiency for HER** was carried out in close accordance with the procedure described in Popczun *et al. J. Am. Chem. Soc.*, 2013, 135, 9267 **:** To quantify the charge to hydrogen formation ratio of the HER, the cathode- and the anode half-cell reactions were separated from each other upon a Nafion® membrane. The distance between RE and WE was adjusted to 1 mm. The distance between RE and CE was ~ 10 mm due to the placement of the membrane in between RE and CE. Hydrogen gas was purged through the cathode half-cell for 3 hours. An inverted solution -containing graduated cylinder was positioned around the working electrode, and the volume of H$_2$ produced whilst chronopotentiometry at j=10 mA cm$^{-2}$ was determined in function of time. The gas amount was compared to the volume calculated from the current passed and the ideal gas law at 293.15 K. The electrode area was 2 cm$^2$. Total duration of the measurement: 4000 s. Volume of the electrolyte: 1700 mL. The results can be taken from

Figure 8a. The red line in Figure 8a corresponds to 100% Faradaic efficiency with a line equation: y=4.9863 x $10^{-3}$x with y=amount of hydrogen ($cm^3$); x=time (s).

**XPS Spectroscopy**

Measurements in Germany (Kuepper, Wollschlaeger):

XPS measurements were performed using a Phoibos HSA 150 hemispherical analyzer equipped with standard Al Kα source with 0.3 eV full width at half-maximum. The measurements were recorded with the sample at room temperature. The spectra were calibrated using the carbon 1s line of adsorbed carbon ($E_B$ = 285.0 eV).

Measurements in Canada (Chevrier, Zhang):

The XPS instrument used for measurements was a Multilab 2000 (ThermoVG Scientific) using a Mg X-ray source for the incident X-rays. Experiments were conducted at room temperature under ultra high vacuum conditions. Samples were etched using an Ar beam to remove adventitious carbon from the surface. Ni42 required 40 minutes of etching and the Ni42-300 required 10 minutes.

**Electron Microscopy**

Cross sectional analysis (vertical plane imaging) of samples was realized by dual beam FIB (focused ion beam) –SEM technique. SEM images of the cross sections were taken on a Zeiss Auriga scanning electron microscope equipped with a Cobra FIB-column and Ga ion source using Feature Milling software module for modeling. The accelerating voltage was adjusted to 15 kV and the SEM images were acquired with a secondary electron detector.

**Gas sorption**

Gas sorption experiments (adsorption/desorption) were carried out using a Quantachrome Autosorb AS-1C. Physisorption was determined at 77 K and 100 K for $N_2$ (Linde Gas, purity > 99.999%) with dynamic $p_0$-determination via a $p_0$-cell at p = 760 mmHg and an Oxford Instruments cryostat model Optistat MK1 equipped with an Oxford Instruments controller model ITC 503. Analyses and interpretation of data were carried out utilizing the Quantachrome AS1Win software package, version 2.11. Prior to the determinations, the samples were activated at 100 °C and pressures of $1\cdot10^{-3}$ mbar for 24h and at $1.5\cdot10^{-6}$ mbar for 24h. All samples were treated in the outgas station until outgassing rates were below 3 microns/minute pressure increase and subsequently contacted with He (Linde Gas, purity > 99.999%) before the analyses were carried out. An equilibration time of 10 min per analysis point was used and ten points were recorded each for adsorption and desorption.

**AFM Experiments**

AFM experiments. AFM measurements were undertaken in semi-contact mode on a NT-MDT model NTEGRA Probe NanoLaboratory. The V-shaped cantilevers had nominal lengths of 140 µm, force constants of 25-95 N/m, and a resonance frequency of 242 kHz (within range of 200-400 kHz). The tip radius was 10 nm. The AFM images were processed using the software Nova Px.

**Determination of nickel and iron ions in the electrolyte after chronopotentiometry**

When pH 7 electrolyte was used Iron ions could form with $PO_4^{3-}$ ions as a precipitation of different species of iron phosphate, but no precipitation at all could be obtained in the electrolyte used for chronopotentiometry.

Twenty mL of the electrolyte used for the long time chronopotentiometry was filled in a 100 mL glass beaker. One mL of a 0,2 M $FeCl_3$ was added and the $FePO_4$ precipitation was filtered. One mL of saturated $H_2O_2$ (30 wt.%, VWR, Darmstadt, Germany) was added to the filtrate and the solution was concentrated by heating it up to 95 °C for 30 min until a total volume of 3 mL was reached. After cooling down to room temperature 1 mL of saturated ammonia (25 wt.%, VWR, Darmstadt, Germany) was added (pH 10). After shaking the mixture for 1 min one mL of a 0.1 M ethanolic Disodium bis- dimethylglyoximate (99% purity, Carl Roth, Karlsruhe, Germany) was added. No red precipitation was formed proving that no Nickel ions were present (detection limit < 1 ppm).

**Determination of nickel and iron ions in the electrolyte used for electro-activation**

20 mL of the electrolyte used for the electro-activation was filled in a 300 mL glass beaker (beaker Nr. 1) and the solution neutralized under stirring by adding around 75 mL of 0.5 M HCl (99% purity, VWR, Darmstadt, Germany) until- pH 6 was reached. An amount of 25 mL was taken from the solution and filled in a second 100 mL glass beaker (beaker Nr. 2).

The solution of beaker Nr. 2 was concentrated upon heating to 95 °C for 30 min until a volume of 3 mL was reached, and after cooling down 60 mg of potassium ferrocyanide $K_4[Fe(CN)_6]$ (99% purity, Carl Roth, Karlsruhe, Germany) were added. The mixture formed was shaken for 1 min. Neither a precipitation nor a colouration took place showing that neither $Fe^{2+}$ nor $Fe^{3+}$ was present (detection limit < 1 ppm). After 20 min 1 mL of saturated $H_2O_2$ (30 wt.%, VWR, Darmstadt, Germany) was added to the remaining solution (beaker Nr.1) and the solution was similarly concentrated by heating to 95 °C for 30 min until the total volume reached 3 mL. After cooling down to room temperature 1 mL of saturated ammonia (25 wt.%, VWR, Darmstadt, Germany) was added (pH 10). No precipitation was formed proving that no iron ions were present. One mL of a 0.1 M ethanolic Disodium bis (dimethylglyoximate) 99% purity; Carl Roth, Karlsruhe, Germany) was added. No red precipitation was formed proving that no Nickel ions were present (detection limit < 1 ppm).

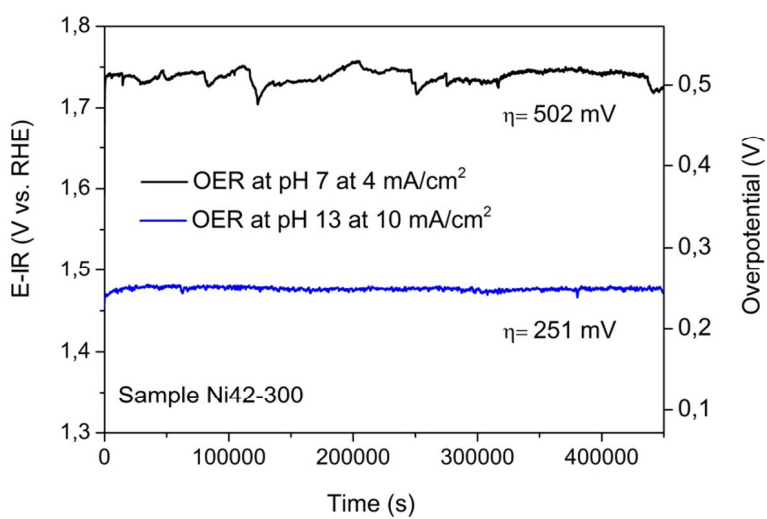

Figure S1. Long term chronopotentiometry measurement of OER upon sample Ni42-300 performed in pH 7 corrected 0.1 M $KH_2PO_4/K_2HPO_4$ at 4 mA/cm² current density (black curve), in 0.1 M KOH at 10 mA/cm² (dashed curve) respectively. Average overpotential for the OER through 450000 s plot: 251 mV (pH 13); 502 mV (pH 7). Electrode area of all samples: 2 cm².

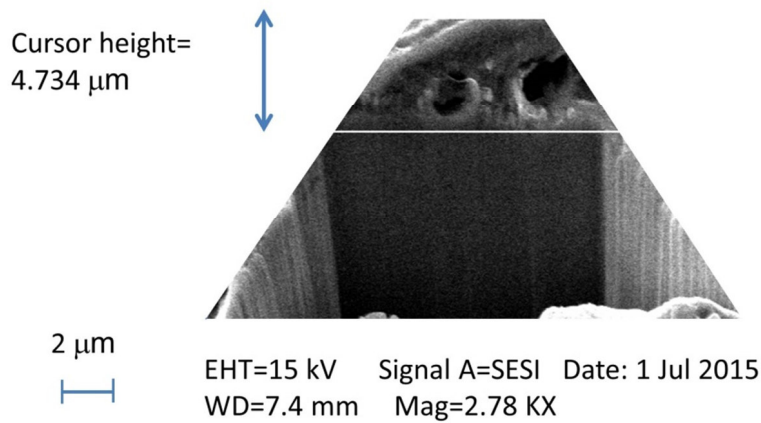

Figure S2. SEM micrograph of a FIB machined cross section. The accelerating voltage was adjusted to 15 kV and the SEM images were acquired with a secondary electron detector; magnification:2780. Ion (Ga) beam settings: current: 600 pA, voltage: 30 kV, duration: 30 min.

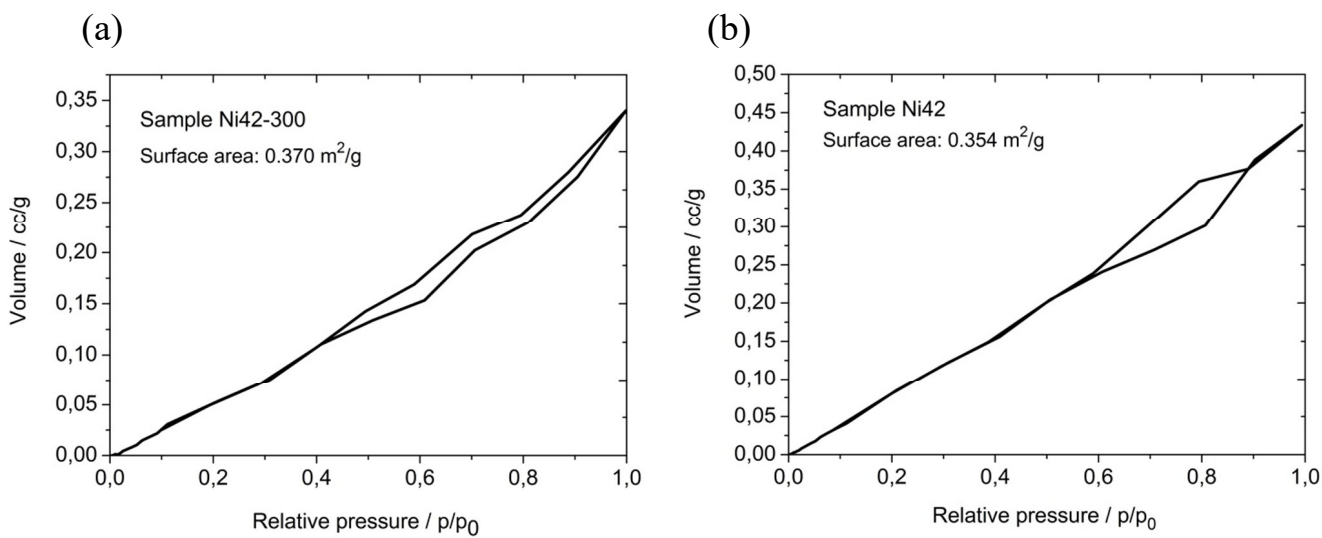

Figure S3. Results from BET measurements. Sample gas: $N_2$ (77K). Activation was performed by outgassing for 24 h at 100 °C ($1*10^{-6}$ bar) and by 24 h at 100 °C ($1.5*10^{-9}$ bar); Equlibrium time: 10 min. (a) Adsorption/desorption plot of sample Ni42-300. (b) Adsorption/desorption plot of sample Ni42.

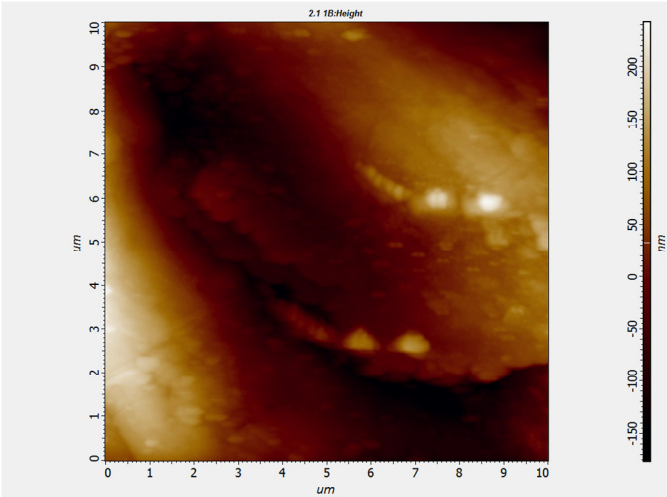

Sampling area: 99.917 um*um  
Height parameters:  
Root mean square roughness: 84.749 nm  
Average roughness: 71.521 nm  
Area peak-to-valley height: 417.145 nm  
Maximum area peak height: 241.657 nm  
Maximum area valley depth: 175.488 nm  
Projected area: 99.917 um*um  
Surface area: 101.778 um*um  
Sample Ni42

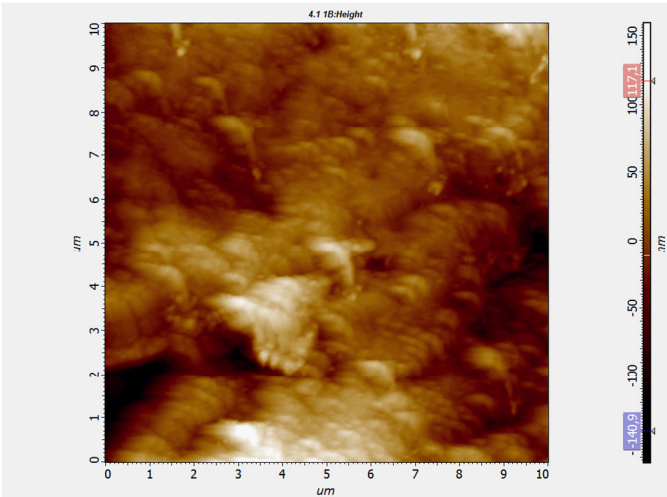

Sampling area: 99.917 um*um  
Height parameters:  
Root mean square roughness: 39.423 nm  
Average roughness: 29.538 nm  
Area peak-to-valley height: 322.02 nm  
Maximum area peak height: 158.734 nm  
Maximum area valley depth: 163.285 nm  
Projected area: 99.917 um*um  
Surface area: 101.797 um*um  
Sample Ni42-300

Figure S4. Images of samples Ni42 and Ni42-300 based on AFM experiments. Analysis of roughness based on the AFM images.

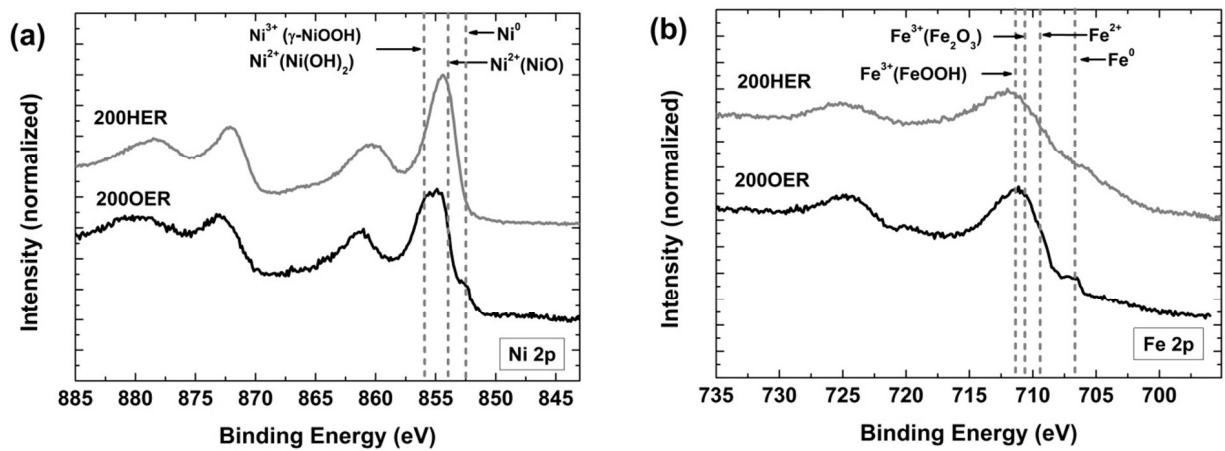

Figure S5. High resolution XPS spectra of sample Ni42-300 recorded after 40000 s of OER upon sample Ni42-300 at 4mA/cm$^2$ in pH 7 solution (200OER), 40000 s of HER at 10mA/cm$^2$ in pH 13 solution (200HER) respectively.

| Sample Activation Process | Mass (g) | Mass difference (g) |
|---|---|---|
| 1 | 5.41199 (5.41203) | -0.00004 |
| 2 | 5.38918 (5.38950) | -0.00032 |
| 3 | 5.42021 (5.42011) | +0.00010 |
| 4 | 5.29711 (5.29741) | -0.00030 |
| 5 | 5.44194 (5.44184) | +0.00010 |

Table S1. Column II: Mass of the steel samples before (in brackets) respectively after carrying out the electro activation procedure. Column III: Mass difference during electro-activation.

| Element | Ni | Fe | C | Position of the 2p$_{(3/2)}$ main lines | |
|---|---|---|---|---|---|
| | | | | Fe | Ni |
| Ni42 | 26.6% | 72.4% | 0.57% | 711.4 eV | 856.0 eV |
| Ni42-300 | 80.8% | 18.55% | 0.67% | 712.0 eV | 855.8 eV (852.4 eV) |

Table S2. Cationic distribution and position of the 2p$_{(3/2)}$ main lines of Fe, Ni and C of samples Ni42 and Ni42-300 derived from the XPS measurements presented in Figure 4.

| Sample | Electrode area(cm$^2$) | Distance between RE and WE (mm) | Distance between RE and CE (mm) | Electrolyte resistance (Ω) at pH 0/1/7/13/14/14.6* |
|---|---|---|---|---|
| Ni42 | 2 | 1 | 4-5 | 0.47/4.1/6.8/5.0/0.6/0.4 |
| Ni42-300 | 2 | 1 | 4-5 | 0.47/4.1/6.8/5.0/0.6/0.4 |
| IrO2-RuO2 | 2 | 1 | 4-5 | 0.47/4.1/6.8/5.0/0.6/0.4 |

Table S3. Key data of the (three electrode) electrochemical set up used for all OER as well as for the cyclic voltammetric- and chronopotentiometry HER measurements at pH 13, 14, 14.6 (0.1 M KOH, 1 M KOH, 7 M KOH), at pH 1, 0 (0.05 H$_2$SO$_4$, 0.5 M H$_2$SO$_4$) and in pH 7 buffer regime (0.1 M KH$_2$PO$_4$/K$_2$HPO$_4$). * The resistances were corrected for 1 mm distance between WE and RE from literature values reported for 10 mm RE-WE distance [114].

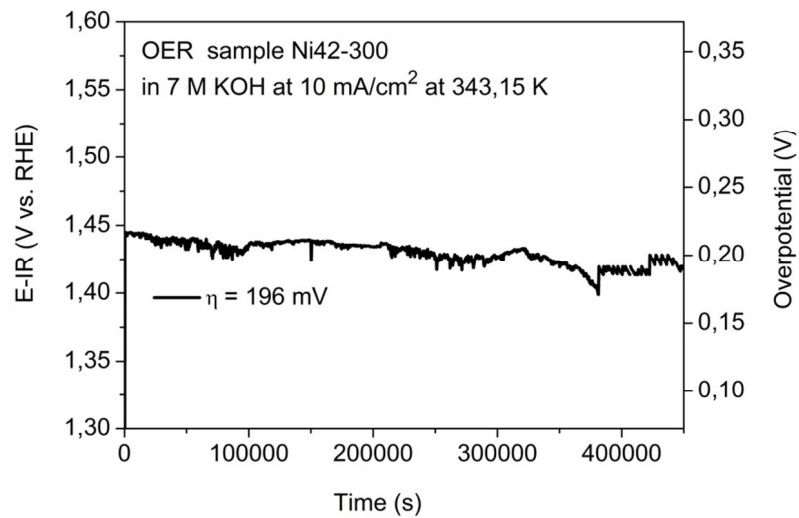

Figure S6. Long term chronopotentiometry measurement of OER upon sample Ni42-300 performed in 7 M KOH at 70 °C. Average overpotential for the OER through 450000 s plot: 196 mV. Electrode area of all samples: 2 cm$^2$.

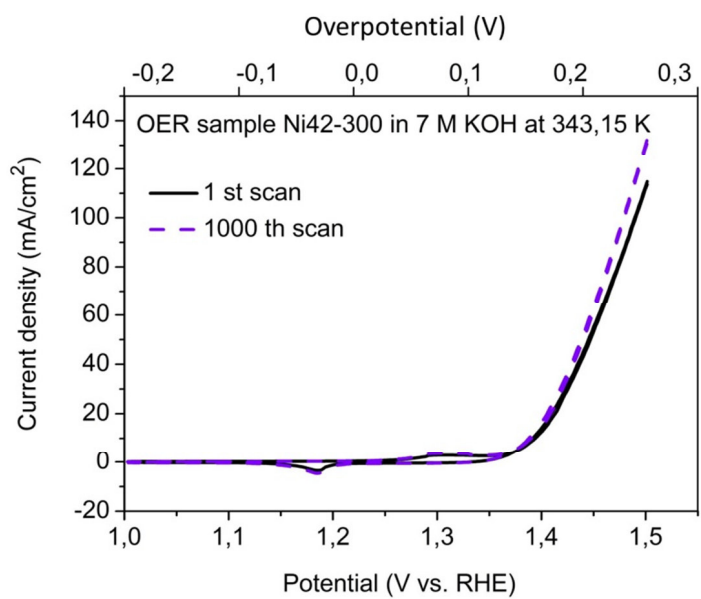

Figure S7. Cyclic voltammogram`s of sample Ni42-300 (first scan: black curve; 1000 th scan: dashed curve) recorded for the HER in 7 M KOH at 70 °C. Settings: 20 mV/s scan rate, 2 mV step size. Electrode area: 2 cm$^2$.

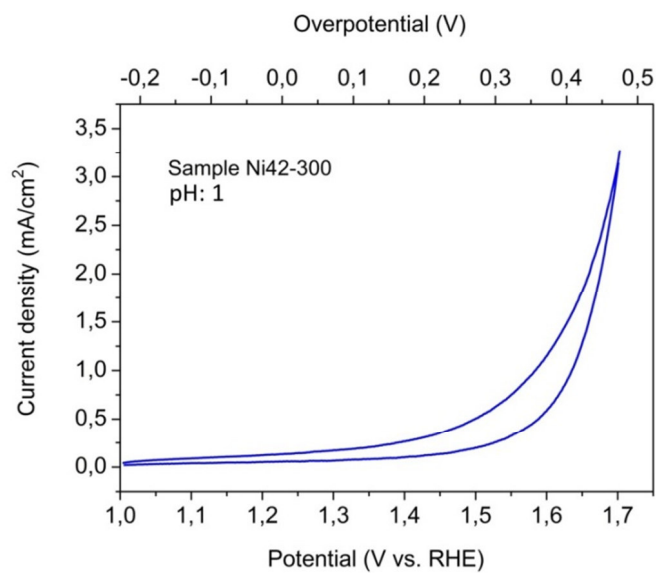

Figure S8. OER performance of sample Ni42-300 at pH 1. The scan rate was set to 20 mV/s.

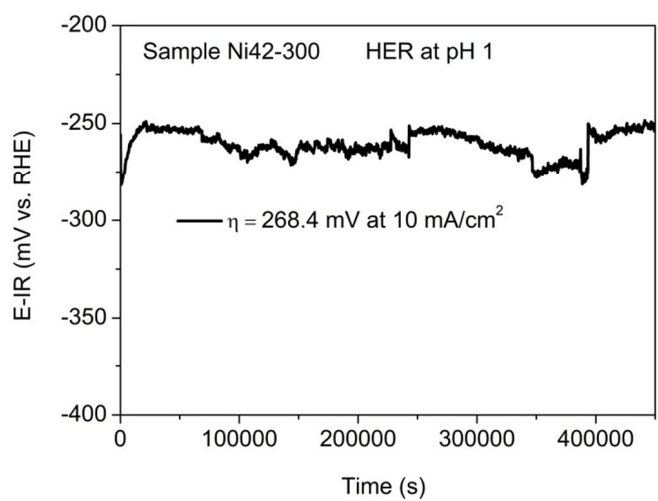

Figure S9. Long term chronopotentiometry measurement of HER upon sample Ni42-300 performed in 0.05 M $H_2SO_4$. Average overpotential for the HER through 450000 s plot: 268.4 mV. Electrode area of the sample: 2 $cm^2$.

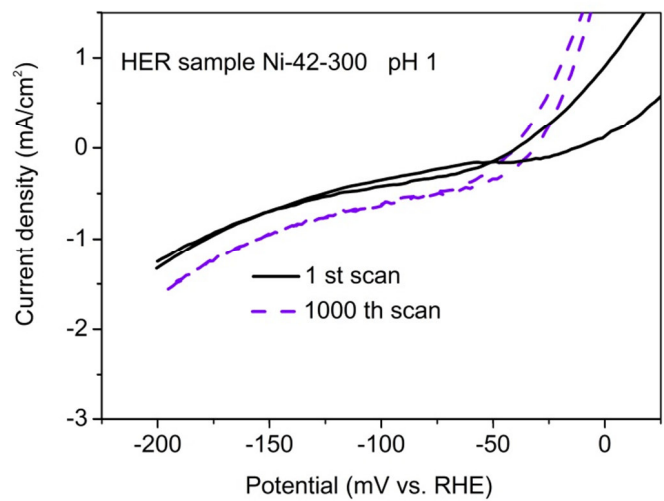

Figure S10. Cyclic voltammogram`s of sample Ni42-300 (first scan: black curve; 1000 th scan: dashed curve) recorded for the HER initiated in 0.05 M $H_2SO_4$. Settings: 5 mV/s scan rate, 2 mV step size. Electrode area: 2 $cm^2$.

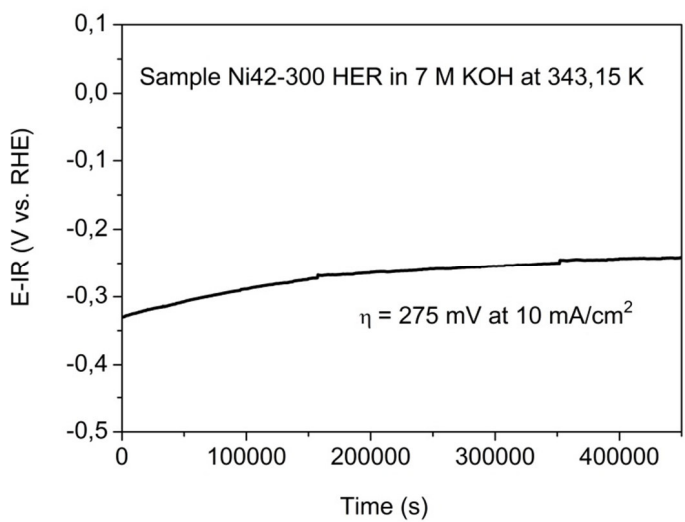

Figure S11. Long term chronopotentiometry measurement of HER upon sample Ni42-300 performed in 7 M KOH at 70 °C. Average overpotential for the HER through 450000 s plot: 275 mV. Electrode area of the sample: 2 $cm^2$.

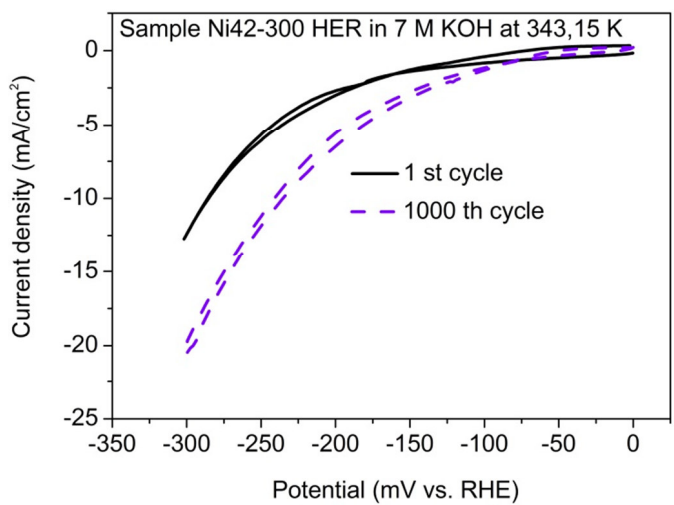

Figure S12. Cyclic voltammogram`s of sample Ni42-300 (first scan: black curve; 1000 th scan: dashed curve) recorded for the HER initiated in 7 M KOH at 70 °C. Settings: 20 mV/s scan rate, 2 mV step size. Electrode area: 2 cm$^2$.

| Material | Overpotential (η) in mV OER (j*, pH) | Overpotential (η) in mV HER (j*, pH) | Tafel slope (mV dec$^{-1}$) OER (pH) | Tafel slope (mV dec$^{-1}$) HER (pH) | Faradaic Eff. (%) OER (pH) | Faradaic Eff. (%) HER (pH) | Ref. |
|---|---|---|---|---|---|---|---|
| Ni42-300 | 491 (4,7) 251 (10,13) 215 (10,14) 196** (10,14.6) | 189 (10,0) 268.4(10,1) 333(10,13) 299(10,14) 275**(10,14.6) | 198.2 (7) 71.6 (13) | 117.9 (0) 80.5(1) 124.4 (13) 117,5(14) | 99.4 (7) | 101.8 (13) | This work |
| Ni$_2$P | 312 (10, 14) | 110 (10,14) | | | 100 (14) | 100 (14) | 103 |
| Co$_2$P | 260 mV (10, 14) | 95 (10,0) | 52 (14) | 45(0) | | 80 (14) | 74 |
| CoP | 320 mV (10, 14) | 89 (10,0) | 64 (14) | 41(0) | | | 74 |
| Fe | 452 (10, 13) | 360 (10, 13) | | | 95.8 (13) | 105.2 (13) | 104 |
| CoSe | 292 (10, 14) | 121 (10, 14) | 69 (14) | 84 (14) | 100 (14) | 100 (14) | 105 |
| NiCo$_2$S$_4$ | 310 (50, 14) | 263 (50, 14) | 89(14) | 141(14) | - | - | 106 |
| NiSe | 317 (100, 14) | 185 (50, 14) | 64 (14) | 120 (14) | 100 (14) | 100 (14) | 107 |

Table S4. Electrochemical OER and HER characteristics of recently developed bi-functional electrocatalysts suitable for full water splitting. The overpotential values are based on long term chronopotentiometry measurements * j= current density in mA/cm$^2$. **Measurement performed at 343.15 k.